\documentclass[twocolumn,aip,jcp,reprint]{revtex4-1}
\usepackage{mathrsfs}
\DeclareMathAlphabet{\mathpzc}{OT1}{pzc}{m}{it}
\usepackage{array}
\usepackage{amsmath}
\usepackage{mathtools}
\usepackage{physics}
\usepackage{amssymb}
\usepackage{graphicx}
\usepackage{subfigure}
\usepackage{color,xcolor}
\usepackage{booktabs}
\usepackage{amstext}
\usepackage{amsfonts}
\usepackage{float}
\usepackage{bm}
\usepackage{epstopdf}
\usepackage{color}
\usepackage{tcolorbox}
\usepackage{orcidlink}
\usepackage{hyperref}
\hypersetup{colorlinks=true
,linkcolor=black
,urlcolor=blue
,citecolor=blue}
\usepackage{dcolumn}
\usepackage[utf8]{inputenc}
\usepackage[T1]{fontenc}
\usepackage{mathptmx}
\usepackage{etoolbox}

\begin{document}

\preprint{AIP/123-QED}

\title{Unveiling the Dynamical Genesis of Quantum Entanglement in Linear Systems: Internal causality breaking in the reduced subsystem evolution}

\author{Shuang-Kai Yang}
\affiliation{Department of Physics and the Center for Quantum
Information Science, National Cheng Kung University, Tainan 70101,
Taiwan}

\author{Wei-Min Zhang }
\email{wzhang@mail.ncku.edu.tw }
\affiliation{Department of Physics
and the Center for Quantum Information Science, National Cheng Kung
University, Tainan 70101, Taiwan}

%\date{\today}

\begin{abstract}
Utilizing the general theory of open quantum systems to investigate the exact dynamical evolution of simple bilinear systems, we discover a mechanism of
the dynamical genesis of quantum entanglement. We focus in detail on the exact quantum evolution dynamics of two photonic modes 
(or any two bosonic modes) coupled to each other through a linear interaction, as the simplest system of open quantum systems that 
we have investigated in the last two decades. Such a linear coupling alone fails to produce two-mode entanglement. We also start 
with an initially separable pure state of the two modes. By solving exactly the quantum equation of motion without relying on the 
probabilistic interpretation, we find that when the initial state of one mode is different from a coherent state (a minimum uncertainty wave 
packet with equal variance in the conjugate quadratures that corresponds to a well-defined classically "particle"), 
the causality in the time-evolution of each mode is internally violated. It also leads to the emergence of quantum entanglement between the two modes. 
The lack of causality is the nature of statistics. We discover that it is the internal violation of causality in the reduced (subsystem) 
dynamical evolution that results in the emergence of entanglement and statistic probability in quantum mechanics, even though 
the dynamical evolution of the whole system completely obeys the deterministic Schr\"{o}dinger equation. This conclusion 
is valid for the quantum dynamics of more complicated composite systems. It may provide the fundamental mechanism of 
the dynamical genesis for both the entanglement and the statistical probability within the deterministic framework of quantum mechanics, 
which is the longest-standing problem that has not been fully understood since the birth of quantum mechanics.
%It also indicates that the probability interpretation introduced in the early development of quantum mechanics is redundant.
\end{abstract}

\maketitle

\section{Introduction}
\label{intro}
Quantum mechanics has been confirmed by countless experiments to be the most powerful theory in the description of 
the natural phenomena.
However, the nature of quantum mechanics itself has been the subject of debate and research since its inception. 
%after Schr\"{o}dinger developed the equation governing the dynamical behavior of matter waves. 
The most controversial issue is about the physical meaning of wavefunctions solved from the deterministic 
Schr\"{o}dinger equation in describing the physical 
quantities measured experimentally in reality. In particular, the Born's  statistical probability interpretation of wavefunctions, which is practically very successful 
for all the observed results in the microscopic world, has put to rest the long historical debate. %begun with the two great physicists, Einstein and Bohr. 
In 1935, Einstein, Podolsky, and Rosen further pointed out that under the probability interpretation, quantum mechanical wavefunctions of distant noninteracting 
systems contain non-local (instant) correlations \cite{einstein1935can}, which goes far beyond the usual understanding of physical observations in reality. 
The non-local correlation feature of wavefunctions in composite systems was soon named as entanglement by Schr\"{o}dinger \cite{schrodinger1935discussion}, 
and becomes the most striking and also the most fundamental property in quantum physics. 

Over the past half century, numerous experiments have been developed to demonstrate the entanglement effects. 
The earliest experimental proof for EPR paradox, which was first recognized by Bohm and Aharonov~\cite{bohm1957discussion}, 
was indeed given by Wu, {\it et al}.~in 1950 in 
their experiment of measuring the angular correlation of scattered annihilation photons \cite{Wu_PR_77}. It further inspired Bell to find mathematically 
an inequality for the correlations between two systems to be satisfied if there are local hidden variables associated with the 
probability description of quantum measurement \cite{bell1964einstein}. Bell also showed that entanglement states of distant noninteracting systems 
could violate this inequality. Thus, violation of Bell's inequality becomes a criterion for demonstrating the non-local property of entanglement.  
Aspect~\cite{aspect1982experimental,aspect1982experimentala}, Clauser~\cite{clauser1969proposed,freedman1972experimental}, 
and Zeilinger~\cite{weihs1998violation} have been awarded the 2022 Nobel Prize in Physics for their groundbreaking experiments with 
entangled photons and pioneering the investigation of quantum information science. 
Nowadays, entanglement has become the most useful resource in the development of new-emerging quantum technologies.

Although the non-local property of entanglement has been well demonstrated by many experiments  based on the violation of Bell's inequality, 
a more fundamental question arisen from EPR paradox is, why and how entanglement emerges in the quantum realm  but not in classical world? 
The non-locality (violation of Bell's inequality) is a sufficient condition for the observation of entanglement between particles at a distant that any 
interaction between them can be ignored. It rules out the possibility of having local hidden variables for quantum probabilistic description. However, it does not 
directly answer the above question. Moreover, within the framework of the Standard Model which is build on the local gauge theory \cite{Weinberg2000}, 
non-local entanglements between various physical systems are all originally generated at an earlier time by the more fundamental local interactions. 
For examples, photon-photon entanglements and electron-electron entanglements are indeed all generated through the local electron-photon 
interactions (or more fundamentally, the electron-photon interaction in QED). To date, there is no definite experimental evidence to show that 
the nature phenomena have gone beyond the predictions from the fundamental local theory, i.e., the Standard Model.  It remains an unsolved 
mystery why entanglement emerges only in the quantum realm but not in classical world.

On the other hand, for any entanglement state of bipartite or multipartite systems, the reduced density matrix of each subsystem 
must be mixed states and cannot be expressed as 
a pure state (wavefunction) that obeys the Schr\"{o}dinger equation. That is, an entanglement state itself is governed 
by Schr\"{o}dinger's deterministic equation of motion but it produces intrinsically the statistical feature for its subsystem states. Apparently, 
the statistical features in entanglement states goes beyond the deterministic framework of the Schr\"{o}dinger equation itself 
because statistical description  is indeterministic, which is a natural consequence of the lack of causality in statistics mechanics.  
In response to the EPR paradox, Bohr had thought  \cite{Bohr1935} that it is the necessity of a final renunciation of the 
classical ideal of causality and a radical revision  of our attitude towards the problem of physical reality. 
In fact, the changes of quantum states are given by two different types 
of interventions, one is the temporally dynamical transformations governed by the system Hamiltonian that are causal (reversible), 
and the other is the abrupt changes brought about by measurements which are non-causal (irreversible) for the system to be observed, 
as described by von Neumann \cite{vonNeumann1932} earlier than the EPR paradox.  To date, the dynamical process of quantum measurements  
is still a fundamental and unsolved problem in modern physics. 

In this paper, without invoking Born’s probabilistic interpretation of wavefunctions or von Neumann’s  ensemble concept  
of quantum measurements or even any other philosophical interpretation in quantum mechanics, we are going to demonstrate 
that a fundamental aspect of quantum entanglement among particles (or subsystems) in composite systems is linked to 
an {\it internal} violation of causality in the reduced
dynamical evolution of each particle (or subsystem).  Thus, the concept of internal causality violation is associated with the 
reduced dynamical evolution of subsystems, which also reflects the system's statistical origin. This concept can be applied to a wide 
range of open quantum systems, from a simple system coupled to a single particle or to a highly complex many-body environment, 
as demonstrated in this study. It shows that despite the fact that the Schrödinger 
equation itself governs the overall evolution of the entire composite systems in a deterministic manner, the internal violation of 
causality in the reduced dynamical evolution for each subsystem occurs as a general consequence of reduced quantum dynamics.
Thus, within the framework of the deterministic Schrödinger equation, a major challenge is to determine under what conditions 
the causal evolution of the subsystems will be destroyed, leading to the emergence of entangled states in the corresponding 
composite system.  
%If we can show this explicitly, 
%it suggests that the probabilistic character of wavefunctions is an intrinsic property of quantum mechanics, not just an 
%assumption, which can potentially resolve the longstanding mystery behind the origin of quantum probability.

To unambiguously address this fundamental issue concerning the dynamical genesis of entanglement and the emergence 
of statistical probability in quantum mechanics, several essential conditions are imposed in this study.
(1) The two sybsystems are initially prepared in separable states, ensuring that no entanglement exists at the outset.
(2) Each subsystem must begin in a pure state rather than a mixed or thermal one, thereby excluding any implicit use of statistical ensembles.
(3) The coupling between the two subsystems is taken to be strictly linear, so that the interaction alone cannot trivially generate entanglement.
(4) The choice of initial separable states is further constrained to be physically realizable, enabling possible experimental verification.
These conditions are designed to eliminate potential ambiguities and loopholes in identifying the genuine dynamical origins of entanglement
and probabilistic nature, beyond the conventional treatments of open quantum systems.

Thus, without loss of the generality, we will investigate the exact quantum dynamics of two photonic modes (or more generally, 
any two bosonic modes) coupled to each other through simple linear exchange interactions, such as those realized with beam 
splitters. Importantly, this type of coupling alone cannot generate two-mode entanglement, as it is well-known in linear quantum 
optics \cite{KLM2001,RB2001}. 
We also begin with an initially separable pure state for the two modes, ensuring that neither entanglement nor any statistical 
correlations are presented at the outset.  
We then solve exactly the deterministic quantum equation of motion through the influence functional \cite{Feynman1963} based 
on the coherent-state path integral in the standard functional formulation of quantum field theory \cite{Faddeev1980}, 
where no a priori statistical probabilistic interpretation is assumed as well. 

In fact, the original Feynman-Vernon influence functional defined in the position representation has been widely used to derive 
the exact master equation of quantum Brownian motion \cite{Leggett1983,GSI1988,HPZ1992,Weissbook}. 
We applied the coherent-state path-integral influence functional to study the reduced dynamics of 
nanoelectronic systems and found the first exact master equation for fermionic systems nearly two decades ago \cite{TZ2008}. 
Around the same time, Prof.~Yijing Yan and his 
collaborators independently employed the influence functional formalism to develop the hierarchical equations of motion 
(HEOM) for describing dissipative dynamics and quantum transport in electronic systems \cite{Jin2008}. This convergence 
of ideas provided an excellent opportunity for collaboration, allowing us to unify two of the most powerful theoretical frameworks: 
the Feynman-Vernon influence functional method and the Schwinger-Keldysh nonequilibrium Green’s function technique 
\cite{Schwinger1961,Keldysh1965,Kadanoff1962}, culminating in a unified description of nonequilibrium reduced quantum  
dynamics \cite{Jin_Tu2010}, which made a significant improvement to the nonequilibrium dynamics combining with the initial-state effect.
Meanwhile, we have continuously extended this framework to derive exact master equations for a broad range of open quantum 
systems, including bosonic, fermionic, and topological systems, by generalizing the formalism within the coherent state 
formulation \cite{LZ2012,LZ2018,Zhang2019}. 

However, in these previous studies on open quantum dynamics, it is typically 
assumed that the environment is initially in an equilibrium state or a system-environment correlated (entangled) state. In both cases, 
the environment is initially in a mixed state, which is inherently a consequence of non-causal statistical average at microscopic level. 
Such assumptions preclude any investigation into the dynamical genesis of entanglement or the emergence of statistical probability, 
since these features are already built into the starting conditions.
In this study, we therefore apply this approach to a "minimal" open system consisting of only one photonic mode interacting 
with another, initially in a separable pure state. One mode is treated as the environment of the other, reacting a minimal 
environment scenario that we will explore in the perspective of open quantum system dynamics. 
Remarkably, we find that the exact master equation for this simplest open quantum system possesses the same formal structure 
as the exact master equations derived for systems coupled to a thermal bath with infinitely many degrees of freedom.
However, their dynamical behaviors are fundamentally different.
The present system is non-dissipative and free of thermal fluctuations, whereas the latter represents a dissipative system governed 
by the well-known fluctuation–dissipation relation that has been extensively studied in open quantum systems.
This clear distinction provides a unique opportunity to connect the deterministic Schrödinger evolution directly with the emergence 
of entanglement and probabilistic behavior in quantum mechanics from first principles, without invoking any statistical assumptions.
The fully solvable model thus serves as an ideal platform for uncovering these phenomena with complete analytical transparency.
%The present work constitutes the application 
%of the same coherent state path-integral formulation to investigate the exact dynamics of a simple non-dissipative system, 
%with the purpose of addressing the fundamental question, namely, the dynamical genesis of entanglement and probability 
%in quantum mechanics.

Because the path integral formulation also provides a direct bridge between quantum 
and classical deterministic dynamics. With the specific setup of the initial states given in the next section, 
it also allows us to show the explicit connection and differences between 
quantum and classical dynamics that neither the Schr\"{o}dinger picture nor Heisenberg picture can demonstrate. 
From such an investigation, we give an answer to the key question explored in this work: why and how entanglement emerges 
only in quantum world but not in classical physics.  Our findings reveal that, unlike classical dynamics, the quantum evolution 
of individual subsystems can violate causality (manifested explicitly how the future dynamics affect the current dynamics
in the reduced dynamical equations of motion). This internal breaking in 
causal behavior is what gives rise to the genesis of entanglement between the two modes, while the two mode system as a whole 
evolves in a causally consistent manner. In other words, the lack of causality in the reduced subsystem evolution underlies the 
statistical, probabilistic nature of quantum states emerging from the deterministic framework of the Schrödinger 
equation.  This finding of the dynamical genesis of entanglement should reveal more profoundly the origin of non-locality 
in quantum mechanics that goes beyond the consequence of Bell’s inequality. The latter only justifies the existence of 
non-locality in quantum mechanics.
%because any measured system cannot be an isolated system due to its interaction with the measurement equipments.  

\section{The formulation of subsystem dynamics using the influence functional method of open quantum systems}
\label{sec2}

%\subsection{Coherent state formulation of Feynman-Vernon influence functional for subsystems}

%To explore the origin of entanglement linked to causality breaking, and to reveal how the statistical characteristics of wave functions 
%emerge from the deterministic quantum equations of motion, we consider a very simple system. 
We begin with a very simple system, a pair of photonic (or more generally, bosonic) modes that are coupled with 
each other through a simple exchange interaction. We treat the dynamics of one mode as an open system 
coupling to the other one as its "environment". The formulation presented below is a simple reduction of the 
general theory of open quantum systems we investigated in the last two decades \cite{TZ2008,Jin_Tu2010,LZ2012,LZ2018}, 
as shown in the end of this section. Explicitly, 
the whole system is described by the following simple Hamiltonian
\begin{align}\label{Htot}
 H_{tot}=\hbar\omega_{1}a_{1}^\dag a_{1}\!+\! \hbar\omega_{2} a_{2}^\dag a_{2}\!+\! \hbar(V_{12}a_{1}^\dag a_{2}\!+\!V^\ast_{12}a_{2}^\dag a_{1}),
\end{align}
where $a_1^\dag$ and $a_2^\dag$ ($a_1$ and $a_2$) are the creation (annihilation) operators of the photonic modes with frequencies
$\omega_1$ and $\omega_2$, respectively. The Hilbert space of the two modes contain infinity number of states (e.g.~in terms of the Fock states
$\{ | n_1, n_2 \rangle = \frac{1}{\sqrt{n_1 ! n_2!} } (a^\dag_1)^{n_1} (a^\dag_2)^{n_2}|00\rangle, n_1, n_2 = 0,1,2,3, \cdots \infty \}$
as the basis). The last term in Eq.~(\ref{Htot}) describes the coupling of the two modes that can be easily realized,  
for example, with a beam splitter in experiments. Because of the linearity, such a coupling by itself does not produce two-mode entanglement \cite{KLM2001,RB2001}. 
We investigate the time evolution of one of the two modes governed by the Hamiltonian in Eq.~(\ref{Htot}). Two distinct initial conditions are considered,
both separable pure states so that no entanglement or statistical features are presented at the beginning.
These are indeed essential requirements to remove the possible loopholes in addressing the nature of quantum states, 
as we have emphasized in the Introduction.  

Although this system is simple 
enough to be solved by anyone who is familiar with quantum mechanics, our application of the path integral technique yields an exact dynamical 
solution that is truly surprising. 
%This result may shed light on a longstanding debate in quantum mechanics regarding the origin of 
%entanglement and the probabilistic interpretation.
Specifically, we let mode 1 be initially in a coherent state (corresponding to a Gaussian wave packet with the minimum Heisenberg 
uncertainty of equal variance in the conjugate quadratures that can serve as a well-defined classical particle), 
the mode 2 can be initially either in a coherent state, or a 
squeezed state (or any other physically reliable pure quantum state). 
As it is well-known, photonic modes are one-to-one corresponding to harmonic oscillators. 
Within the framework of Schr\"{o}dinger's deterministic equation of motion, the time evolution of the coherent states for a single harmonic oscillator  
mode moves exactly along the trajectories of an isolated classical harmonic oscillator for all kind of initial coherent states. This was originally 
discovered by Schr\"{o}dinger in 1926 \cite{Sch_1926_WP} and later
 developed by Glauber for quantum optics in 1963 \cite{glauber1963coherent}, also see the review article by one of the authors 
 \cite{ZhangRMP1990} and the recently published book  \cite{Zhangbook2023}. 
%As it is well-known, Schr\"{o}dinger 's original wave packet for harmonic oscillators or Glauber's coherent state for photons is a quantum state 
%that owns all properties of describing the motion of a classical harmonic oscillator. 
%This quantum state is commonly called the harmonic oscillator coherent state or Glauber coherent state in the literature.
Thus, if mode 1 does not couple to mode 2, the quantum state will move exactly along the classical trajectories. 
 On the other hand, squeezed states are generally the minimum Schr\"{o}dinger uncertainty wave packets that are often 
 used for noise reduction in one of the quadratures \cite{Stoler1970,Yuen1976,Squstate,Schmuc1980,ZhangRMP1990}.
 %Even when their conjugate variances exceed the standard Heisenberg product due to covariance, squeezed states still 
 %saturate the Robertson–Schrödinger uncertainty bound \cite{}, marking them as irreducibly quantum.
They exhibit noise in one quadrature below the vacuum (standard quantum) limit, which is not possible in classical theory. 
Thus, different from the coherent state, squeezed states are intrinsically nonclassical Gaussian states. 
 They provide one of the most direct tests of quantumness in continuous-variable systems. 
Its observation serves simultaneously as a witness of nonclassicality and a benchmark for accessing the quantum regime.

With the above setting, we are going to show explicitly in what conditions the quantum dynamics of the coupled two modes
can give rise to exactly the same classical dynamics. And, under what circumstances the quantum dynamics will 
be deviated away  from the classical dynamics such that the two modes evolve into an entangled state.
We find that the time evolutions of the coupled two modes governed by the Hamiltonian of Eq.~(\ref{Htot}) with the two different initial 
 states mentioned above behave very differently. The detailed calculations are given in Sec.~\ref{sec3}-\ref{sec4}. Simply speaking, 
 for the first case, the quantum coherent state (corresponding to a well-defined classical particle) of each mode keeps in a pure state 
at any later time,  because the coupling Hamiltonian of Eq.~(\ref{Htot}) alone cannot generate entanglement between the 
two modes. The corresponding quantum dynamical evolution follows exactly the classical trajectory solution, as one expected.
While, for the second case, the two modes will eventually entangled together, due to the initial squeezed state (or any other quantum state). 
Such entanglement has indeed been demonstrated in the quantum photonic circuit experiments \cite{QIC, Paesani2019,Arrazola2021}. 
As a result, the state of each mode becomes a mixed state in the second case, the statistical feature and the entanglement 
of two modes naturally emerge.  

These results demonstrate why classical physics cannot have entanglement. Entanglement generated from the 
deterministic equation of motion must accompany with the emergence of statistical probability for its subsystems, 
as a consequence of quantum measurements. 
This suggests that a key principle inherent to classical deterministic dynamics must be violated during the quantum 
evolution in some way when entanglement emerges. We further identify that this is the principle of causality that is
broken down in the {\it internal} dynamical evolution of reduced subsystems. 
Here the word "{\it internal}" has the specific meanings: 
it describes an intermediate process of events, particularly that the process happens between its constitutes or subsystems 
within a larger closed system, it is hidden from direct view and is a necessary step that is not the final output. We will show in detail 
how this happens in the rest of the work.
Such kind of causality breaking may also give a direct answer to the key issue arisen from the EPR paradox, 
namely, quantum world would emerge entanglement but classical physics does not. This answer is obtained from the 
exact solution of reduced quantum dynamics, rather than a kind of interpretation from Bell's inequality
or any other quantum interpretation. To show explicitly our finding, we solve the reduced quantum dynamics in
a very unique way  \cite{LZ2012,Zhang2019}.
%This formulation can make an unambiguous quantum-to-classical correspondence \cite{ZhangPR95}. 
We then analytically and exactly obtain the quantum dynamics of the coupled two modes governed by Eq.~(\ref{Htot}), 
and identify explicitly where and how a breakdown in causality occurs in the reduced dynamical evolution. 

Explicitly,  the dynamics of the two modes are governed by the Schr\"{o}dinger equation,
\begin{subequations}
\begin{align}
   i \hbar \frac{d}{dt} |\psi_{tot}(t)\rangle =H_{tot} |\psi_{tot}(t)\rangle , \label{Sche}
\end{align}
or equivalently  by the  von Neumann equation \cite{vonNeumann1932}
\begin{align}
\dfrac{d}{dt}\rho_{tot}(t)=\dfrac{1}{i\hbar}[H_{tot},\rho_{tot}(t)] , \label{vNe}
\end{align}
\end{subequations}
where $\rho_{tot}(t)= |\psi_{tot}(t)\rangle \langle \psi_{tot}(t)|$ is the density matrix operator of the two modes for the pure total state 
$|\psi_{tot}(t)\rangle$. If $|\psi_{tot}(t)\rangle$ is an 
entangled state, then the reduced density matrix operator $\rho_1(t)= \Tr_2[ \rho_{tot}(t)]$ or $\rho_2(t)= \Tr_1 [\rho_{tot}(t)]$ must be a mixed 
state, namely $\rho^2_1(t)\neq \rho_1(t)$ and $\rho^2_2(t)\neq \rho_2(t)$. Thus, it is more convenient to 
begin with the von Neumann equation (\ref{vNe}) using the density matrix operator. Starting with wavefunctions and the Schr\"{o}dinger equation is 
inconvenient when we study the reduced dynamics of subsystems in which mixed states will naturally occur \cite{vonNeumann1927,Landau1927}. 
%we use the path integral technique in the coherent state representation \cite{ZhangRMP1990} with some extensions \cite{TZ2008,LZ2012,LZ2018}. 
%We will use the Feynman-Vernon influence functional method for open quantum systems to solve exactly the reduced density matrix 
%operator $\rho_1(t)$ from Eq.~(\ref{vNe}), and to demonstrate how the entanglement emerges, in terms of the same language used in the description 
%of the deterministic classical dynamics. This formulation can make a unambiguous quantum-to-classical correspondence \cite{ZhangPR95}.

To be more specific, the reduced dynamics of the subsystem is given by $\rho_1(t)$, which can be obtained by partially tracing over 
all the states of mode 2 from the total density operator of the 
two modes. The formal solution of Eq.~(\ref{vNe}) for the total density matrix operator is formally determined by \cite{vonNeumann1932}
\begin{subequations}
\label{unitary_evolution}
\begin{align}
    \rho_{tot}(t)=U(t,t_0)\rho_{tot}(t_0)U^\dag(t,t_0)   \label{ted}
\end{align}
with the time-evolution operator
\begin{align}
    U(t,t_0)=\exp\Big[-\frac{i}{\hbar}H_{tot}(t-t_0)\Big].
\end{align}
\end{subequations}
Then, the dynamics of the subsystem $\rho_1(t)$ is given by
\begin{align}
\rho_1(t)=\Tr_2[ U(t,t_0)\rho_{tot}(t_0)U^\dag(t,t_0)]   \label{red}
\end{align}
Because of the coupling between the two modes, as shown in Eq.~(\ref{Htot}), the partial trace in Eq.~(\ref{red})
is not easy to carry out. 

For initially separable states of the two modes, we can easily compute this partial trace by integrated out exactly all 
possible paths for mode 2 in the coherent-state path-integral formulation  \cite{Faddeev1980,ZhangRMP1990}. 
To do so, we express the coherent state matrix element of  $\rho_1(t)$ in terms of the coherent-state path integrals  
as follows \cite{TZ2008,Jin_Tu2010,LZ2012}
\begin{align}
\label{rho_1(t)}
    \langle z_{1f} |\rho_1(t)|z'_{1f} \rangle
     % &= \langle z_{2f}|\Tr_2(\rho_{tot}(t))|z'_{2f}\rangle  \notag\\
    & = \!\! \int \! d\mu(z_{1i}) \mathcal{J} (z_{1f},z'_{1f},t;z_{1i},z'_{1i},t_0)  \notag\\
    & \qquad    \times \! \langle z_{1i}|\rho_1(t_0)|z'_{1i}\rangle d\mu(z'_{1i}),
\end{align}
where the propagating function
$\mathcal{J}(z_{1f},z'_{1f},t; z_{1i},z'_{1i},t_0)$ is defined by
\begin{align}
\label{propagating}
\mathcal{J}(z_{1f}, & z'_{1f},t;  ~z_{1i},z'_{1i},t_0) \notag \\
     \equiv \!\!  \int &d \mu(z_{2f}) d\mu(z_{2i})d\mu(z'_{2i}) \langle z_{1f}z_{2f}|U(t,t_0)|z_{1i}z_{2i}\rangle  \notag\\
   &\! \times  \! \langle z_{2i}|\rho_2(t_0)|{z'_{2i}}\rangle \langle z'_{1i}z'_{2i}|U^\dag(t,t_0)|z'_{1f}z_{2f}\rangle \notag  \\
   % e^{(-\abs{z_{1i}}^2-\abs{z'_{1i}}^2-\abs{z_{1f}}^2)} \notag\\
    = B & ( z_1,  z'_1) \!\! \int \!
    e^{\frac{i}{\hbar} (S_1[z_1]-S_1^\ast[z'_1])}  F[z_1,z'_1]\mathcal{D}[z_1]\mathcal{D}[z'_1].
\end{align}
Furthermore, all the dynamical effect of mode 2 on the mode 1 can encompassed into the following path-integral influence functional in Eq.~(\ref{propagating}),
which was originally introduced by Feynman and Vernon \cite{Feynman1963} but was extended and derived in the coherent-state path-integral formulation \cite{LZ2012,Zhang2019},
\begin{align}
\label{influence}
%    \begin{split}
F[z_1,z'_1]& = \!\! \int \! d\mu(z_{2f}) d\mu(z_{2i})d\mu({z'_{2i}})  
  \langle z_{2i}|\rho_2(t_0)|{z'_{2i}}\rangle  \notag \\
&\! \times B(z_2,z'_2)\!\! \int \! e^{\frac{i}{\hbar}
(S_{2_1}[z_1,z_2]-S^\ast_{2_1}[z'_1,z'_2])} \mathcal{D}[z_2]\mathcal{D}[z'_2].
%\end{split}
\end{align}

In Eq.~(\ref{propagating}), the matrix $\langle z_{1f}z_{2f}|U(t,t_0)|z_{1i}z_{2i}\rangle$ describes the forward propagating function 
and  $\langle z'_{1i}z'_{2i}|U^\dag(t,t_0)|z'_{1f}z_{2f}\rangle$ the backward propagating for the whole system, they are Hermitian conjugates 
to each other, representing the causal and time symmetric evolution of the Schr\"{o}dinger equation for the closed system.
The factors in Eqs.~(\ref{propagating})-(\ref{influence}) 
The factor $ B(z_j,z'_j)=\exp(\Phi(z_j)+\Phi^\ast(z'_j))$ is the boundary effect in the coherent-state path-integral formulation
with $\Phi(z_j)=\frac{1}{2}{z^\ast_{jf}} z_j(t)+\frac{1}{2}z^\ast_j(t_0)z_{ji}$ for $j=1,2$ specifying mode 1 and mode 2. 
%with $\Phi(z_j)=\frac{1}{2}{z^\ast_{jf}} z_j(t)+\frac{1}{2}z^\ast_j(t_0)z_{ji}$ for $j=1,2$ specifying mode 1 and mode 2. 
This boundary factor was originally discovered by Feddeev and Slavnov in formulating the functional quantum field theory with the 
coherent-state path-integral  formulation \cite{Faddeev1980}. 
We should emphasize here that these boundary factors are crucially 
important in the coherent-state path integrals. It gives the deterministic transition amplitudes between the two coherent states that 
does not presented in the oirginal Feynman's path integral formalism. It dominates the most important contribution to the influence 
functional and the propagating functional of the reduced density matrix, as we will show later.

The functions $S_1[z_1]$, $S_{2_1}[z_1,z_2]$ in Eqs.~(\ref{propagating})-(\ref{influence}) are the classical actions of the two 
coupled photonic modes in the coherent state representation, coming straightforwardly  from the forward time-evolution operator 
$U(t,t_0)$ in coherent-state path integrals,
\begin{subequations}
\label{csaction}
\begin{align}
%\begin{cases}
    S_1[z_1]=\!\! \int^t_{t_0} \!\!\Big(\frac{i\hbar}{2}(z^\ast_1\dot{z}_1-\dot{z}_1^\ast z_1)-\hbar\omega_1z_1^\ast z_1\Big)d\tau ,   \label{action_1} \\
    S_{2_1}[z_1,z_2]= \!\!\int^t_{t_0} \!\! \Big(\frac{i\hbar}{2}(z^\ast_2\dot{z}_2-\dot{z}_2^\ast z_2)-\hbar\omega_2z_2^\ast z_2 \notag \\
    - \hbar V_{12}z^\ast_1z_2-\hbar V_{12}^\ast z_2^\ast z_1\Big)d\tau .   \label{action_2}
 %   \mathcal{L}_{12}[z_1,z_2]=\mathcal{L}_1[z_1]-\mathcal{H}_{12}[z_1,z_2]
%\end{cases}
\end{align}
\end{subequations}
The backward time-evolution operator $U^\dag(t,t_0)$ contributes the path integrals with the complex conjugate 
actions $S^*_1[z'_1]$ and $S^*_{2_1}[z'_1,z'_2]$ in Eqs.~(\ref{propagating}-\ref{influence}).
Here, to avoid the unambiguous boundary conditions in coherent state path integral formalism 
\cite{ZhangRMP1990,Schulman1981,Faddeev1980},
we have used the non-normalized coherent state $|z_j \rangle=\exp(z_ja^\dag_j )|0\rangle, j=1,2$ with the resolution 
of identity $\int |z_j\rangle \langle z_j| d\mu(z_j)=1$, where $d\mu(z_j)=e^{-\abs{z_j}^2}\frac{dz^\ast_j dz_j}{2\pi i}$. 
The coherent-state path integral measure is given by $\mathcal{D}[z_j]=\prod_{t_0<\tau<t}\frac{dz^\ast(\tau) dz(\tau)}{2\pi i}$. 

Thus, we complete the formulation  of the reduced density dynamics $\rho_1(t)$ in terms of the coherent-state path-integral 
influence functional for such a simple coupled two-mode system for our further investigation.
This formulation is indeed a simple reduction of the general influence functional 
theory for open quantum systems in which the environment is usually consisting of an infinite number of degrees of freedom 
\cite{LZ2012,Zhang2019}. It can be seen clearer if we extend the subsystem of the second mode to be
a larger system containing an infinite number of modes with different frequencies, namely, extending Eq.~(\ref{Htot}) to the 
following one:
\begin{align}
\label{Hetot}
 H^{\rm ext}_{tot}= & \hbar\omega_{1}a_{1}^\dag a_{1}\!+\! \sum_k \hbar\omega_{k} a_{k}^\dag a_{k}
  +\! \sum_k \hbar(V_{1k}a_{1}^\dag  a_{k}\!+\!V^\ast_{1k}a_{k}^\dag a_{1}),
\end{align}
where $a_k^\dag$ ($a_k$) is the creation (annihilation) operators of the photonic modes with frequency
$\omega_k$, and the summation of $k$ contains an infinite number of modes. Then, the above formulation remains the same
except that the coherent-state influence functional of Eq.~(\ref{influence}) needs to include the contributions from all other modes \cite{LZ2012}:
\begin{align}
\label{ext_influence}
%    \begin{split}
F^{\rm ext}[z_1, & z'_1] = \!\! \prod_k \!\! \int \! d\mu(z_{kf}) d\mu(z_{ki})d\mu({z'_{ki}})  
  \langle z_{ki}|\rho_k(t_0)|{z'_{ki}}\rangle  \notag \\
&\! \times B(z_k,z'_k)\!\! \int \! e^{\frac{i}{\hbar}
(S_{k_1}[z_1,z_k]-S^\ast_{k_1}[z'_1,z'_k])} \mathcal{D}[z_k]\mathcal{D}[z'_k].
%\end{split}
\end{align}
where $S_{k_1}[z_1,z_k]$ has the same form as Eq.~(\ref{action_2}) by the replacement of the index $2$ by $k$.
This is the general influence functional path integral formulation for open quantum systems in the coherent state 
representation \cite{LZ2012,LZ2018}.

%This is because a coupling to subsystem containing discrete number of degrees of freedom 
%only shifts each mode in the spectra (shift all the normal modes) in both the system and the  such that particles 
%remain oscillational transitions quantum mechanically between the system and the “bath”, no dissipation occurs. 
%For a real bath with continuous spectra, the shift of normal modes is no longer possible because of the continuousness 
%of spectra. As a result, the particle oscillational transitions between the system and the bath are modified by 
%(dissipative) damping motions. In other words, dissipation dynamics of the system can only be caused by 
%continuous spectra of a bath. This is a Many researchers without carefully and explicitly working on the exact 
%dynamics of open quantum systems probably cannot recognize such important consequence for the dissipative 
%quantum dynamics in open quantum systems.

In this paper, we reduce the environment to be a very simplified one that contains only one mode, as shown by the 
Hamiltonian of Eq.~(\ref{Htot}), and focus on the reduced dynamical dynamics of the subsystem from the perspective of open quantum systems. 
%As it is well-known the dynamics of an open quantum system is very different from that of a closed system. 
%The former involves non-unitary evolutions, while the latter undergoes unitary evolutions. 
Directly solving the Schr\"{o}dinger equation 
of Eq.~(\ref{Htot}) may not be easy to see how the unitary evolution is broken down for the reduced dynamical evolution of subsystem.
Intuitively, it is difficult to imagine how the evolution of subsystem can become non-unitary in such a simple coupled two-mode system.  
However, when all dynamical paths of mode 2 are integrated out using the path-integral 
influence functional given in the above equations, 
it naturally mixes the forward and backward propagating paths of mode 1 through the simple 
exchange transitions between the two modes of Eq.~(\ref{Htot}), it is this mixture breaks not only the unitary evolution but also
the causality of subsystem evolution. This is a general and inherent property of the reduced quantum dynamics.
In the next section, we will show in detail how the exact dynamical evolution of the subsystem 
(individual mode) becomes {\it internally} non-unitary and non-causal, even though the whole system of two modes together 
follows the deterministic unitary evolution and is causal.

\section{Solving the exact influence functional path integrals with stationary paths for the subsystem dynamics}
\label{sec3}

\subsection{Exact path integrals in bilinear systems}

Path integral is defined as a sum over all possible paths (quantum-mechanically) for the system evolving from one state to 
another \cite{Feynman_Hibbs1965,Schulman1981}. However, 
stationary paths alone are sufficient to represent the path integral of Eqs.~(\ref{propagating}-\ref{influence}) in such 
a bilinear system.  
In the coherent-state path-integral formulation \cite{Faddeev1980}, this becomes analytical transparency because the path 
integral is directly given by stationary paths satisfying the classical equations of motion in the complex phase 
space\cite{TZ2008,Jin_Tu2010,LZ2012,LZ2018,LZ2018,Zhang2019}, see a simple illustration given in Ref.~[\cite{Illustration}]. 
It provides a specific way to solve exactly the quantum dynamics directly with classical mechanics.
%It provides a specific way  how to solve exactly the quantum dynamical directly with classical equations of motion. 
Explicitly, the stationary paths obey the classical equations of motion determined by the least action principle 
$\delta S= \delta \!\int_{t_0}^t \mathcal{L} [z_i(\tau), z^*_i(\tau)]d\tau =0$. In other word, the stationary paths satisfy the
Euler-Lagrange equation,
\begin{align}\label{EL-eq(F)}
    \frac{d}{d\tau}\frac{\partial\mathcal{L}} {\partial \dot{z}_i}-\frac{\partial \mathcal{L}}{\partial z_i}=0~,
    ~~ \frac{d}{d\tau}\frac{\partial\mathcal{L}} {\partial \dot{z}^*_i} - \frac{\partial \mathcal{L}}{\partial z^*_i}=0 .   
\end{align}
%This allows us to solve the quantum mechanics in terms of classical dynamics. 
%Thus, it also provides a unambigorous connection 
%between the quantum and classical dynamics that can help us to answer the questions
%discussed in the introduction in such a simple bilinear system.
From Eq.~(\ref{action_2}), we obtain the classical equations of motion for mode 2,
\begin{subequations}
\label{eom-z_2,z_2*}
\begin{align}
     \frac{d z_2(\tau)}{d\tau}+i\omega_2 z_2(\tau)+iV^\ast_{12}z_1(\tau)=0 ,  \label{m2z}\\
     \frac{d z^\ast_2(\tau)}{d\tau}-i\omega_2z^\ast_2(\tau)-iV_{12}z^\ast_1(\tau)=0  . \label{m2zs}
\end{align}   
\end{subequations}

It is easy to check that with the one-to-one transformation maps the coherent-state complex space to the usual 
phase space \cite{ZhangRMP1990},  
\begin{align}
z_2=\frac{1}{\sqrt{2}} (x_2+ip_2)~,~~ z^*_2=\frac{1}{\sqrt{2}} (x_2-ip_2), \label{correspd}
\end{align}
where $x_2$ and $p_2$ represent the dimensionless position of momentum of the mode as a harmonic oscillator
(the dimension factors are omitted for the simplicity of discussions), Eq.~(\ref{eom-z_2,z_2*}) gives the exact classical 
equation of motion for mode 2 of the two coupled harmonic oscillators in the familiar phase space. Since 
no a priori probabilistic interpretation is assumed in the coherent state path integrals, the stationary paths solved from 
above equations of motion are the same as the classical trajectories of harmonic oscillator coupled to another harmonic oscillator. 
On the other hand, Eq.~(\ref{m2zs}) is the conjugation of Eq.~(\ref{m2z}). This is also a basic mathematical consequence that quantum evolution 
must obey the complex structure arisen from the unitary evolution that ensures the time symmetry in quantum evolution, i.e.~the reversibility, 
just as the mathematical requirement of the symplectic structure for classical evolution of conservative systems \cite{ZhangPR95,Helgason1978}. 
Thus, quantum and classical evolution dynamics are unified in the same framework in the coherent-state path-integral formulation. 

%In general, “paths” in Feynman’s path integral are usually 
%not physical trajectories that the particle fellows, only the entire sum over paths gives the transition amplitudes that makes the 
%measurable predictions \cite{man_Hibbs1965,Schulman1981}. However, for linear systems, the entire sum over paths is represented
%by the stationary paths, as we discussed in the beginning of this section. Meantime, we choose the initial state of the system 
%(mode 1) in a coherent state, which is a non-spreading wave packet with minimum uncertainty, its evolution alone precisely fellows 
%the classical trajectories.

Not only these two equations of motion are conjugated to each other, their boundary conditions are also conjugated, i.e., the boundary 
conditions of Eqs.~(\ref{m2zs}) and (\ref{m2z}) are given by $z_2(t_0)=z_{2i}$ and $z_2^\ast(t)=z_{2f}^\ast$, respectively
 \cite{LZ2012}. This is also a natural quantum mechanics result, namely, one cannot fixed the boundary condition 
of $z$ and $z^*$ (or equivalently the position and the momentum) at the same time, due to the uncertainty relationship.
As a result, the solutions of these two equations of motion in Eq.~(\ref{eom-z_2,z_2*}) are given by, 
%and these two equations (Eq.~(\ref{sol-z1,z1*})) mean that two independent variables $z_1(\tau)$, $z_1^\ast(\tau)$ evolve respectively 
%in the forward time-direction and in the backward time-direction.
\begin{subequations}
\label{sol-z2,z2*}
\begin{align}
    &z_2(\tau)=z_{2i}e^{-i\omega_2(\tau-t_0)}-iV^\ast_{12} \!\! \int^{\tau}_{t_0} \!\!\! e^{-i\omega_2(\tau-\tau')}z_1(\tau') d\tau' , \\
    &z_2^\ast(\tau)=z_{2f}^\ast e^{-i\omega_2(t-\tau)}-iV_{12} \!\! \int^{t}_{\tau} \!\!\!e^{-i\omega_2(\tau'-\tau)}z^\ast_1(\tau') d\tau' ,
\end{align}
\end{subequations}
where $t_0 \le \tau \le t$, which agrees with the condition for the unitary evolution. 
Equation (\ref{sol-z2,z2*}) describes the full dynamics of mode 2 for the forward quantum evolution in Eq.~(\ref{influence}). 
%The stationary paths of  Eq.~(\ref{sol-z2,z2*}) are exactly the same as the classical trajectories of harmonic oscillator with the same coupling to another
%harmonic oscillator in complex phase space. 
Take a special case that let two modes decouple, i.e. $V_{12}=0$, the solution of Eq.~(\ref{sol-z2,z2*}) is simply 
reduced to the solution obtained exactly from the quantum evolution for a single harmonic mode, as shown in the illustration 
\cite{Illustration}. It demonstrates how the stationary path uniquely determines the coherent-state path integral.
Likewise, we can find the stationary paths of mode 2 for the backward quantum evolution, i.e.~the
solution of the variables $z'_{2}(\tau)$ and ${z'_2}^\ast(\tau)$ from the stationary path equation of motion,
% and use the backward evolution matrix element in propagating function (Eq.~(\ref{propagating})) to find the boundary condition. 
%Then, we can get the following equations (Eq.~(\ref{sol-z1',z1'*})), and these two equations (Eq.~(\ref{sol-z1',z1'*})) mean that two 
%independent variables $z'_1(\tau)$, ${z'_1}^\ast(\tau)$ evolve respectively in the backward time-direction and in the forward time-direction.
\begin{subequations}\label{sol-z2',z2'*}
    \begin{align} &z'_2(\tau)=z_{2f}e^{i\omega_2(t-\tau)}+iV^\ast_{12} \!\! \int^{t}_{\tau} \!\!\! e^{i\omega_2(\tau'-\tau)}z_1(\tau') d\tau' , \\
    &z'^\ast_2(\tau)=z'^\ast_{2i} e^{i\omega_2(\tau-t_0)}+iV_{12} \!\! \int^{\tau}_{t_0} \!\!\! e^{i\omega_2(\tau-\tau')} z'^\ast_1(\tau') d\tau'.
\end{align}
\end{subequations}
Thus, the path integrals for mode 2 given in the influence functional of Eq.~(\ref{influence}) is fully determined by the stationary paths given by  
Eqs.~(\ref{sol-z2,z2*}) and (\ref{sol-z2',z2'*}).
%So far, the quantum and classical solutions are one-to-one correspondence.

Remarkably, when we substitute the above solutions into Eq.~(\ref{influence}) and completely integrate out the degrees of freedom of mode 2,
 the effect of mode 2 on the dynamics of mode 1 could be very different for different initial states of mode 2.  
To see the exact dynamical effect of mode 2 on mode 1, namely, to calculate the rest integrals in Eq.~(\ref{influence}), we need
to specify the initial states $|\psi_{tot}(t_0)\rangle$ of the total system. As we have discussed in Sec.~\ref{sec2}, we shall choose a separable initial state of 
the two modes so that no entanglement begins with:
\begin{align}
|\psi_{tot}(t_0)\rangle =  |\alpha_1\rangle \otimes |\alpha_2, s \rangle ,  \label{initials}
\end{align}
where $ |\alpha_1\rangle=D(\alpha_1)|0\rangle)$ is the Glauber coherent state and  $|\alpha_2, s \rangle=D(\alpha_2)S(s)|0\rangle$ is a 
squeezed coherent state. The displacement operator $D(\alpha_i)=\exp(\alpha_i a^\dag_i-\alpha^\ast_i a_i)$ with ($i=1,2)$ and the squeezed 
operator $S(s)=\exp(\frac{1}{2}(s {a^\dag_2}^2-s^\ast a^2_2))$. 
The coherent parameters $\alpha_i$ and the squeezing parameter $s$ are complex numbers. 
We may rewrite the squeezing parameter as $s=\gamma e^{i\theta}$. 
Because Eq.~(\ref{initials}) is a pure state, it is a vector in the Hilbert space of the two modes. There is also no statistic feature to 
begin with it if one does not artificially impose the quantum state a probability interpretation.
Thus, the two special initial state setups mentioned in Sec.~\ref{sec2}  correspond to $s=0$ (i.e. $\gamma=0$) and $\alpha_2=0$, respectively. 
%{\color{red}In other words, we put the initial state at the boundary betw}
%Namely,  in the first initial state, two modes are both in coherent states, while in the second initial state, mode 1 is in a coherent state and mode 2 is in a squeezed state.   

Now, we go to show explicitly how the two different initial states of the mode 2 result in very different reduced dynamical evolution for the subsystem
of mode 1. In terms of the density operator, the initial states of the mode 2 in Eq.~(\ref{initials}) can be written as
%\begin{subequations}  
\begin{align}\label{ini_rho}
&   \rho_2(t_0)=  |\alpha_2, s \rangle \langle \alpha_2, s |.
\end{align}
%\end{subequations}
Thus, the influence functional of Eq.~(\ref{influence}) can be explicitly and exactly calculated. With the help of the faithful representation of the generalized Heisenberg group 
$H_6$ (a subgroup of the symplectic Lie group $Sp(4)$  \cite{ZhangRMP1990}) plus Gaussian integrals, it is easy to find the influence functional of Eq.~(\ref{influence}),
\begin{align} 
F[z_1,z'_1]=\exp(\frac{i}{\hbar} S_{\rm eff}[z_1,z'_1]).  \label{IFs}
\end{align}
Here, $S_{\rm eff}[z_1,z'_1]$ denotes an effective action describing the dynamical effect arisen from mode 2 on mode 1 through their coupling Hamiltonian in 
Eq.~(\ref{Htot}). It is given by
% the The influence Lagrangian  $\mathcal{L}_{IF}[z_1,z'_1]$ is given by
\begin{subequations}
\label{IFes}
\begin{align}
%\begin{split}
     S_{\rm eff}[z_1,  z'_1]&  =   S_{\rm co}[z_1, z'_1] + S_{\rm sq}[z_1, z'_1]  \notag \\
    &+ i\hbar \! \int^t_{t_0} \!\! d\tau \!\! \int^\tau_{t_0} \!\! \!d\tau'  \big[\chi^{*}_1({\tau})g(\tau,\tau')z_1(\tau') \notag \\
   &%\!\!-i\hbar\!\int^\tau_{t_0}
   \qquad \qquad  \qquad -\! {z'_1}^\ast(\tau')g(\tau',\tau)\chi^{ }_1({\tau})\big]   ,
%\end{split}
\end{align}
where 
%\begin{widetext}
\begin{align}
%\begin{split}
    S_{\rm co}[z_1, z'_1]=&\!-\!\hbar V^\ast_{12} \alpha^\ast_2 \!\! \int^t_{t_0} \!\! d\tau e^{i\omega_2(\tau-t_0)}\chi^{ }_1({\tau}) + \text{c.c.} , \label{IFco} \\
    S_{\rm sq}[z_1, z'_1] =& \!-\!i\hbar \!\int^t_{t_0} \!\! d\tau \!\! \int^t_{t_0} \!\!\! d\tau'  \Big\{ 
    \chi^{*}_1({\tau}) \widetilde{g}(\tau,\tau')\chi^{ }_1({\tau'})  \notag \\
   & \quad - \frac{1}{2}[ \chi^{*}_1({\tau}) \overline{g}(\tau,\tau')\chi^{*}_1({\tau'}) + c.c.]
%    \frac{\tanh\gamma e^{i\theta}}{1-\tanh^2\gamma} \left[\left(\frac{1}{2}(V_{12})^2e^{-i\omega_2(\tau-t_0)}\chi^\ast(\tau)\right.\right.\\
 %   \qquad \cross\left.\left.\!\!\int^t_{t_0}\!\!e^{-i\omega_2(\tau'-t_0)}\chi^\ast(\tau')d\tau'\right.+\text{c.c.}\right) .
    \Big\} , \label{IFsq}
%\end{split}
\end{align}
%\end{widetext}
\end{subequations}
are the contributions arisen from the coherent part and the squeezing part of the initial state in  Eq.~(\ref{ini_rho}), respectively.
In Eq.~(\ref{IFes}), we have introduced the variable $\chi^{ }_1({\tau'}) = z^{ }_1(\tau') \!-\! {z'_1}(\tau')$ which characterizes the
difference between the forward and the backward paths of the density matrix evolution. 
%Here, the variable $\chi(\tau)$ is $\chi(\tau)=z_1(\tau)-z'_1(\tau)$, 
Also, there are three two-time correlation functions in Eq.~(\ref{IFes}), which are given by
\begin{subequations}
\label{ttcf}
\begin{align}
g(\tau,\tau') &=  \abs{V_{12}}^2e^{-i\omega_2(\tau-\tau')},  \\
\widetilde{g}(\tau,\tau') &= % \frac{\tanh\gamma^2}{1-\tanh^2\gamma} 
\sinh^2\!\gamma \abs{V_{12}}^2e^{-i\omega_2(\tau-\tau')}, \\
\overline{g}(\tau,\tau') & = \frac{\sinh(2\gamma)e^{i(\theta+2\omega_2t_0)}}{4} 
% \frac{\tanh \gamma\cdot e^{i(\theta+2\omega_2t_0)}}{1-\tanh^2\gamma}  
 (V_{12})^2e^{-i\omega_2(\tau+\tau')},
\end{align}
\end{subequations}
These correlation functions characterize the non-local two-time correlations between the two modes and between the forward and 
the backward propagating paths. 
The physical meaning of  Eq.~(\ref{IFes}) is clear:  the coherence part in the initial state of mode 2 acts as a linear driving field applying on mode 1, see Eq.~(\ref{IFco}).
While the squeezing part induces non-local time correlations between the forward and backward evolutions for the reduced density matrix of 
 mode 1 through the coupling with mode 2 at different times, see Eq.~(\ref{IFsq}), which are indeed the sources of the causality violation in the 
 dynamical evolution of mode 1 as well as the origin of the entanglement emergence between the two modes. 
 It is particularly important to 
 notice that the influence functional (an exact coarse-graining) contains all possible dynamical influence of mode 2 (or the environment) on mode 1
(the system) but it did not tell how the detailed reduced dynamics of the system behaves.  To fully understand the dynamical behaviors of
the reduced system, one should solve exactly the path integral of the propagating functional in Eq.~(\ref{propagating}) for the reduced 
density matrix, which is a {\it fine-graining} that is crucial for understanding the reduced quantum dynamics. 
In the next subsection, we provide such a fine-graining computation.

 \subsection{The internal causality breaking of the stationary paths for each individual mode}
Having the exact analytical solution of the influence functional of Eq.~(\ref{influence}), given by Eqs.~(\ref{IFs})-(\ref{IFes}), 
we can now solve the propagating function Eq.~(\ref{propagating}) to determine the time-evolution of mode 1. Substituting Eq.~(\ref{IFes}) into Eq.~(\ref{propagating}), the classical action of the mode 1 in the path integrals 
is modified as $S_1[z_1]- S^\ast_1[z'_1]+S_{\rm eff}[z_1, z'_1]$, where $S_1[z_1]$ is the classical action of mode 1 for the forward evolution, 
$S^\ast_1[z'_1]$ is that of the backward evolution, and $S_{\rm eff}[z_1, z'_1]$ given in Eq.~(\ref{IFs}) is the effective action induced 
by mode 2 on mode 1 through the two-mode coupling in Eq.~(\ref{Htot}), which mixes the forward and backward paths together. 
Note that the total action for mode 1 is still a quadratic function 
of the complex variables $z_1$ and $z'_1$. Thus, the path integrals of Eq.~(\ref{propagating}) are again fully determined by the stationary paths 
%of the complex variables $z_1$, $z'_1$, $z^\ast_1$ and ${z'_1}^\ast$ 
which obey the Euler-Lagrange equation governed by the action $S_1[z_1]-S^\ast_1[z'_1]
+S_{\rm eff}[z_1, z'_1]$. The resulting equations of motion of mode 1 for the forward path $(z_i(\tau), z^*_i(\tau))$ and the backward 
path $(z'^*_i(\tau), z'_i(\tau))$are
\begin{subequations}
\label{eom-1}
\begin{align}
&\dot{z}_1(\tau)\!+\!i\omega_1z_1(\tau) \!+\!\!\!\int^\tau_{t_0}\!\!\!\!  g(\tau,\tau')z_1(\tau')d\tau'=\!V_{12}(\tau)\alpha_2 \notag \\
 & \qquad \!+\!\!\int^t_{t_0}\!\!\! \big[\overline{g}(\tau,\tau')\chi^{*}_1({\tau'})
 \!-\! \widetilde{g}(\tau,\tau') \chi^{ }_1({\tau'}) \big] d\tau', \label{eom-z1}  \\
%\end{align}
%\begin{align}
&\dot{z}_1^\ast(\tau)\!-\!i\omega_1z_1^\ast(\tau)\!-\!\!\!\int^t_\tau\!\!\! g^\ast(\tau,\tau')z_1^\ast(\tau')d\tau'\!=\!V^\ast_{12}(\tau)\alpha_2^\ast\! \notag \\
&\!\!-\!\!\int^t_{t_0}\!\!\!\! \big[g^\ast\!(\tau,\tau'){z'_1}^\ast\!(\tau')\!+\!\overline{g}^\ast\!(\tau,\tau')\chi^{ }_1\!({\tau'})\!-\!\widetilde{g}^\ast\!(\tau,\tau')\chi^{*}_1({\tau'}) \!\big]\!d\tau'\! , \label{eom-z1*}  \\
%\end{align}
%\begin{align}
 &{\dot{z}_1}'^\ast \!(\tau) \!-\! i\omega_1{z'_1}^\ast \! (\tau) \!+\!\!\!\int^\tau_{t_0}\!\!\!\! g^\ast(\tau,\tau'){z'_1}^\ast \! (\tau')d\tau' \! =V^\ast_{12}(\tau)\alpha_2^\ast   \notag \\
 & \qquad -\!\!\int^t_{t_0}\!\!\! \big[\overline{g}^\ast(\tau,\tau')\chi^{ }_1({\tau'})  \!-\! \widetilde{g}^\ast(\tau,\tau')\chi^{*}_1({\tau'})\big] d\tau' , \label{eom-z'1*}  \\
% \end{align}
%\begin{align}
   &\dot{z}'_1(\tau)+i\omega_1z'_1(\tau)\!- \! \!\! \int^t_\tau \!\! \! g(\tau,\tau')z'_1(\tau')d\tau' \! = \!V_{12}(\tau)\alpha_2  \notag \\
    &\! - \!\!  \int^t_{t_0} \!\! \! \big[ g(\tau,\tau')z_1(\tau') \! - \! \overline{g}(\tau,\tau')\chi^{*}_1({\tau'}) 
     \!+\!\widetilde{g}(\tau,\tau')\chi^{ }_1({\tau'})\big] d\tau'  ,  \label{eom-z'1} 
\end{align}
\end{subequations}
subjected to the boundary conditions $z_1(t_0)=z_{1i}$, $z^\ast_1(t)=z^\ast_{1f}$,  $z'^*_1(t_0)=z'^*_{1i}$ and $z'_1(t)=z'_{1f}$, 
where $t_0 \le \tau \le t$ and $V_{12}(\tau)=-iV_{12}e^{-i\omega_2(\tau-t_0)}$. 

Because $\chi^{ }_1({\tau'}) = z^{ }_1(\tau') \!-\! {z'_1}(\tau')$, Eq.~(\ref{eom-1}) shows that the forward and backward 
paths are all mixed together. 
Combining the solutions of Eq.~(\ref{eom-1}) with the solutions of Eqs.~(\ref{sol-z2,z2*})-(\ref{sol-z2',z2'*}) 
for mode 2, it shows that the forward and backward paths of mode 2 are mixed as well. 
As one can see, Eqs.~(\ref{eom-z1}) and (\ref{eom-z1*}) [also Eqs.~(\ref{eom-z'1*}) and (\ref{eom-z'1})] are no longer 
conjugated each other. In other words, we start with a unitary evolution formulation for the two-mode coupling system but
the unitarity is broken when we look at the dynamical evolution of each individual mode. Here, breaking unitary
simply means that the equations of motion for the variable $z_1(\tau)$ and $z^*_1(\tau)$ are no longer conjugated each 
other. In fact, breaking unitary is indeed a general consequence for open quantum systems, namely for any open quantum system, 
%composite quantum system (a system consists of two or more coupled subsystems, up to an infinity number of subsystems), 
its dynamical evolution must be non-unitary because the openness of the system can cause dissipation and fluctuations
which cannot be unitary.  %\cite{TZ2008,LZ2012,LZ2018,Zhang2019}. 
The above equations of motion indicates further that for any finite composite system (consisting of even only two subsystems 
or two particles, where there exists no dissipation as we will shown later), 
the quantum dynamical evolution of each subsystem also cannot be unitary as long as they coupled each other. 

%Before we go to solve the stationary paths and the reduced density matrix $\rho_1(t)$ for mode 1, we must point out  a very 
%important consequence arisen from the equations of motion 
Moreover, these equations of motion in Eq.~(\ref{eom-1}), which determine all
the contributions for the path integrals in Eq.~(\ref{propagating}),
show that not only the quantum unitary evolution is broken, 
the causality of its dynamical evolution is also broken for each mode, even though these equations are derived from the
deterministic evolution equation of Eq.~(\ref{unitary_evolution}) without taking any approximation. Note that $t_0 \le \tau \le t$, Eq.~(\ref{eom-z1})
shows that the solution of the forward path $z_1(\tau)$ at time $\tau$ evolves from $t_0$ to $\tau$,
but it is also affected by its future dynamics from $\tau$ to $t$, see the last term in Eq.~(\ref{eom-z1}).  
Explicitly, the first two terms in the first line of Eq.~(\ref{eom-z1}) is the dynamics of mode 1 itself. The third term is an integral 
from $t_0$ to $\tau$ with $g(\tau,\tau')$ as its integral kernel which describes the forward dynamical processes of the two-mode 
coupling, and the term in the right side of equality is an equivalent driving force induced by the coherent part of the initial state
of mode 2. These terms preserve the causality of the evolution of mode 1. But the second line in Eq.~(\ref{eom-z1}) is an integral from 
$t_0$ to $t ~(> \tau) $ with integral kernels $\widetilde{g}(\tau,\tau')$ and $\overline{g}(\tau,\tau')$ which are proportional to the quadratures
${\rm tr}[a^\dag_2 a_2 \rho_2(t_0)]-|{\rm tr}[a^\dag_2 \rho_2(t_0)] |^2=\sinh^2\gamma$ and ${\rm tr}[a_2 a_2 \rho_2(t_0)]=\sinh2\gamma
e^{i\theta}/4$, respectively. 
The proportionality to $\chi^{ }_1({\tau'})$ also mixes the forward and the backward evolutions together in these equations of motion.
This integral consists of not only the past evolution contribution from $t_0$ to $\tau$ but also the 
advanced (future) evolution contribution from $\tau$ to $t$. It is the latter contribution violates the causality in terms of the language 
of classical deterministic dynamics.    
Likewise, Eq.~(\ref{eom-z'1}) also shows that the backward path at time $\tau$ evolves 
from $t$ to $\tau$, but it is also influenced by its "future" dynamics from $\tau$ to $t_0$.  
The other two equations of motion, i.e., Eqs.~(\ref{eom-z1*}) and (\ref{eom-z'1*})  for their conjugate variables, 
show the same property of the violation of the causality.
%and meantimes also show the breaking of the unitary, as we have already pointed out.

In general, causality breaking is mathematically defined as follows: for an equation of motion written as 
$\dot{x}(t)=F(x(t-\tau),x(t),x(t+\sigma)...)$, causality is broken whenever there exists a $\sigma >0$ such that 
$\partial F/\partial x(t+\sigma) \neq 0$. That is, its evolution at time $t$ depends on dynamical variables 
evaluated at future times. 
%Thus, we can give a mathematical definition of {\bf causality breaking}: a non-stochastic equation of motion contains terms 
%from the future dynamics that change the present dynamics of the system.  
Thus, the stationary path equations of motion of Eq.~(\ref{eom-1}) shows explicitly the causality breaking. It demonstrates 
physically and mathematically how both the past and the future dynamics of the subsystem change its present dynamics.
This is for the first time in the literature showing how a dynamical equation of motion derived from the deterministic evolution
manifests explicitly the violation of causality.
It is a rigorously derived mathematical consequence rather than a philosophical argument. 
It shows that the violation of the causality only occurs for the dynamics of individual mode rather than the dynamics of 
two modes as a whole. Therefore, we called such kind of causality violation as {\bf internal causality breaking}, 
to distinguish the causal evolution of the Schr\"{o}dinger equation or the von Neumann equation for the isolated system.
Here the word "{\bf Internal}" becomes clear: it represents the intermediate processes of reduced 
subsystem evolution between the initial time $t_0$ and a later time $t$. These processes happen between 
subsystems. Thus, internal causality breaking is given by the mixing dynamical effects non-locally between the 
two modes (or more general, between the reduced system and the rest part in the closed system), and also the mixing dynamical effects 
non-locally between the forward and backward propagating paths of mode 1 (the subsystem). This is a profound  
discovery that reveals the origin of non-locality in quantum mechanics. It has gone beyond the consequence of 
Bell’s inequality which only justifies the existence of non-locality in quantum mechanics without giving the physical reason. 

%Our concept of the internal causality breaking describes precisely violation of causality in , it is hidden from direct view 
%and is a necessary step that is not the final output, as it will becomes clearer in the next discussions.}

Due to the internal causality breaking, finding the corresponding solutions of these equations of motion is 
usually not an easy task.
However,  using the approach we have developed in solving the dynamics of open quantum systems in the last two decades \cite{TZ2008,Jin_Tu2010,LZ2012,LZ2018,Zhang2019}, the mixed forward and backward paths with the causality breaking can actually be solved analytically. 
This is done by introducing a formal solution (a linear transformations) to Eq.~(\ref{eom-1}), 
\begin{subequations}
\label{z1 relation}
\begin{align}  
    &z_1(\tau)=  u(\tau,t_0)z_{1i}+v_0(\tau,t_0)\alpha_2 \notag \\
    & \qquad \quad -v_1(\tau,t)\chi(t)-v_2(\tau,t)\chi^*(t),   \label{sz1}\\
    & {z'_1}^\ast(\tau)=  u^\ast(\tau,t_0){z'}^\ast_{1i}+v_0^\ast(\tau,t_0)\alpha^\ast_2\notag \\
    & \qquad \qquad +v^\ast_1(\tau,t)\chi^*(t)+v^\ast_2(\tau,t)\chi(t), \\
    & \chi(\tau) =  u^\ast(t,\tau)\chi(t), ~~\chi^*(\tau) =  u(t,\tau)\chi^*(t)
\end{align}
\end{subequations}
where $\chi(t)=z_1(t)-z'_{1f}$ and $\chi^*(t)=z^\ast_{1f}-{z'}^\ast_1(t)$, and $z_1(t)$ and ${z'}^\ast_1(t)$ are the end point values 
of the stationary paths that can be self-consistently determined from the above solution, while $z'_{1f}$ and $z^\ast_{1f}$ are the fixed
end point values in formulating the path integrals, see Eq.~(\ref{rho_1(t)}).

Using the above transformation, the equations of motion Eq.~(\ref{eom-1}) are reduced to
\begin{subequations}
\label{eom-evo}
\begin{align}
&\dot{u}(\tau,t_0)+i\omega_1u(\tau,t_0)+\!\int^\tau_{t_0}\!\! g(\tau,\tau')u(\tau',t_0)d\tau'\!=\!0, \\
&\dot{v}_0(\tau,t_0)+i\omega_1v_0(\tau,t_0)+\!\int^\tau_{t_0}\!\!g(\tau,t')v_0(t',t_0)dt' = V_{12}(\tau) , \\
&\dot{v}_1(\tau,t)+i\omega_1v_1(\tau,t)+\!\int^\tau_{t_0}\!\!g(\tau,\tau')v_1(\tau',t)d\tau'\notag\\
&~~~~~~~~~~~~~~~~~~~~~~~~~~~~~~~~~~~~~~~=\!\int^t_{t_0}\!\!\widetilde{g}(\tau,\tau')u^*(t,\tau')d\tau' , \label{eom-v1} \\
&\dot{v}_2(\tau,t)+i\omega_1v_2(\tau,t)+\!\int^\tau_{t_0}\!\!g(\tau,\tau')v_2(\tau',t)d\tau'\notag.  \\
&~~~~~~~~~~~~~~~~~~~~~~~~~~~~~~~~~~~~~~~=\!-\!\int^t_{t_0}\!\!\overline{g}(\tau,\tau')u(t,\tau')d\tau' ,  \label{eom-v2}
\end{align}
\end{subequations}
subjected to the boundary conditions: $u(t_0,t_0)=1$, $v_0(t_0,t_0)=0$, $v_1(t_0,t)=0$ and $v_2(t_0,t)=0$. 
These time-correlation functions correspond indeed to the nonequilibrium Green's functions in quantum 
many-body systems, as we have shown in
solving the general reduced dynamics of open quantum systems \cite{TZ2008,Jin_Tu2010,LZ2012,LZ2018,Zhang2019}. 
%Here it shows that  our general theory for open quantum systems is also applicable to a simple coupled two-mode systems.
To distinguish these nonequilibrium Green's functions in the reduced quantum dynamics from the standard Schwinger-Keldysh Green's 
functions  in many-body systems, we name them as the reduced nonequilibrium Green's functions.
The property of causality breaking is now transformed to the deterministic equations of motion for the reduced Green's functions $v_1(\tau,t)$ 
and $v_2(\tau,t)$.  Explicitly, the inhomogeneous term in Eqs.~(\ref{eom-v1})-(\ref{eom-v2}) shows how the future dynamics change the present 
dynamics. Note that these reduced correlated Green's functions characterizing the correlations in the reduced system in the different times can be directly measured in
experiments. Thus, the internal causality breaking is not only defined mathematically but can also be justified experimentally. 
The nonequilibrium Green's function formalism of Eq.~(\ref{eom-evo}) is universally applicable for many-body interacting systems
and quantum field theory \cite{CSHY1985,RS1986,WMZ1992}. This indicates that internal causality breaking
is a universal quantum property in quantum dynamics in the perspective of open quantum systems.

The analytical solution of these reduced nonequilibrium Green's functions for this simple two-mode coupled system can be easily obtained:
%\begin{subequations}
%\label{eom-1r}
%\begin{align}
%& \dot{u}^\ast(t,\tau)\!+\!i\omega_1u^\ast(t,\tau) \!-\!\!\!\int^t_{\tau}\!\!\!\!  g(\tau,\tau')u^\ast(t,\tau')d\tau'=0 \\
%& \dot{u}(t,\tau)\!-\!i\omega_1u(t,\tau) \!-\!\!\!\int^t_{\tau}\!\!\!\!  g^*(\tau,\tau')u(t,\tau')d\tau'=0 \\
%\end{align}
%\end{subequations}
\begin{subequations}
\label{s_negf}
\begin{align}
     u(\tau,t_0) & =\cos^2(\frac{\varphi}{2})e^{-i\omega_{+}(\tau-t_0)}+\sin^2(\frac{\varphi}{2})e^{-i\omega_-(\tau-t_0)} , \label{uts} \\
     v_0(\tau,t_0) & =-iV_{12}\int^\tau_{t_0} \!\! u(\tau,\tau_1)e^{-i\omega_2(\tau_1-t_0)}d\tau_1  ,  \\
     v_1(\tau,t) & =\! \int^\tau_{t_0}  \!\!d\tau_1 \int^t_{t_0}  \!\! d\tau_2  u(\tau,\tau_1)\widetilde{g}(\tau_1,\tau_2) u^\ast(t,\tau_2) ,  
   \notag \\   & = \sinh^2\gamma v_0(\tau,t_0) v^*_0(t,t_0)   
     \label{v1s}  \\
    %    & \qquad \qquad \qquad \qquad \cdot\!\int^t_{t_0}u^\ast(t,\tau_2)e^{i\omega_2(t'-t_0)}dt'\\
    v_2(\tau,t) & = -\!\! \int^\tau_{t_0}  \!\!d\tau_1 \int^t_{t_0}  \!\! d\tau_2  u(\tau,\tau_1)\overline{g}(\tau_1,\tau_2) u(t,\tau_2) 
   \notag \\ &  = \frac{1}{4}\sinh(2\gamma) e^{i\theta} v_0(\tau,t_0) v_0(t,t_0),  
  \label{v2s} 
    %e^{i\theta}\frac{\tanh\gamma}{1-\tanh^2\gamma}(V_{12})^2\int^\tau_{t_0}u(\tau,t')e^{-i\omega_2(t'-t_0)}dt'\\
   % &~~~~~~~~~~~~\cdot\!\int^t_{t_0}u(t,t')e^{-i\omega_2(t'-t_0)}dt' 
\end{align}
\end{subequations}
where $\omega_{\pm}=\frac{1}{2}(\omega_1+\omega_2)\pm\frac{1}{2}\sqrt{[(\omega_1-\omega_2)]^2+4\abs{V_{12}}^2}$ and $\varphi=\tan^{-1}(\frac{2\abs{V_{12}}} {\omega_1-\omega_2})$. 
Equations (\ref{z1 relation}) to (\ref{s_negf}) give the exact analytical solution of the stationary paths determined by the equations of motion 
of Eq.~(\ref{eom-1}). These analytical solutions make the internal causality breaking more transparent. 
For example, the solution of the forward path 
$z_1(\tau)$ given by Eq.~(\ref{sz1}) contains four terms. The first two terms [$\sim u(\tau,t_0)$ and $v_0(\tau,t_0)$] 
(corresponding to the reduced retarded Green's function and the external field driving Green's function, respectively) are arisen from its past 
historical motion with the coupling to the dynamics of another mode and remain causal. 
The last two terms [$\sim v_1(\tau,t)$ and $v_2(\tau,t)$] (corresponding to the reduced correlated Green's 
function and the reduced pairing correlated Green's function) are contributed from its forward stationary paths mixing with the 
backward stationary paths. It is the latter two reduced correlated Green's functions break the internal causality in the reduced dynamics of the subsystem.
It occurs only if the squeezing parameter in the initial state of the mode 2 does not vanish, i.e., $s \neq0$ in Eq.~(\ref{initials}).
These two reduced correlated Green's functions vanish in any classical deterministic evolution because of the causality, as we will show next.

If $s= 0$, i.e.~$\gamma= 0$ so that both modes are initially in coherent states, see Eq.~(\ref{initials}), then the two-time correlations 
$\widetilde{g}(\tau,\tau') =0$ and $\overline{g}(\tau,\tau') =0$, see 
Eq.~(\ref{ttcf}). This directly leads to $v_1(\tau,t)=0$ and $v_2(\tau,t)=0$, as shown by Eqs.~(\ref{v1s}) and (\ref{v2s}). 
As a result, Eq.~(\ref{z1 relation}) is reduced to 
\begin{subequations}
\label{clas}
\begin{align}
&z_1(\tau)=  u(\tau,t_0)z_{1i}+v_0(\tau,t_0)\alpha_2, \\
&{z'_1}^\ast(\tau)=  u^\ast(\tau,t_0){z'}^\ast_{1i}+v_0^\ast(\tau,t_0)\alpha^\ast_2 ,
\end{align} 
\end{subequations}
which shows that the causality is recovered. 
In other words, if the initial states of both modes are coherent states, they correspond to wave packets with the minimum uncertainty 
$\Delta x = \Delta p$ and $ \Delta x \Delta p= 1/2 $. In this situation, the causality maintains in quantum dynamical evolution for each mode 
in this coupled two-mode system. Note that the wave packets with minimum Heisenberg uncertainty have been described and defined 
as classical particles (classical harmonic oscillators). 
The solution of Eq.~(\ref{clas}) fully agrees with the causal classical solution, 
namely, the quantum evolution of coherent states can reproduce the exact causal classical dynamics of each mode in this two coupled harmonic oscillators.  

\begin{figure}
\includegraphics[width=0.9 \linewidth]{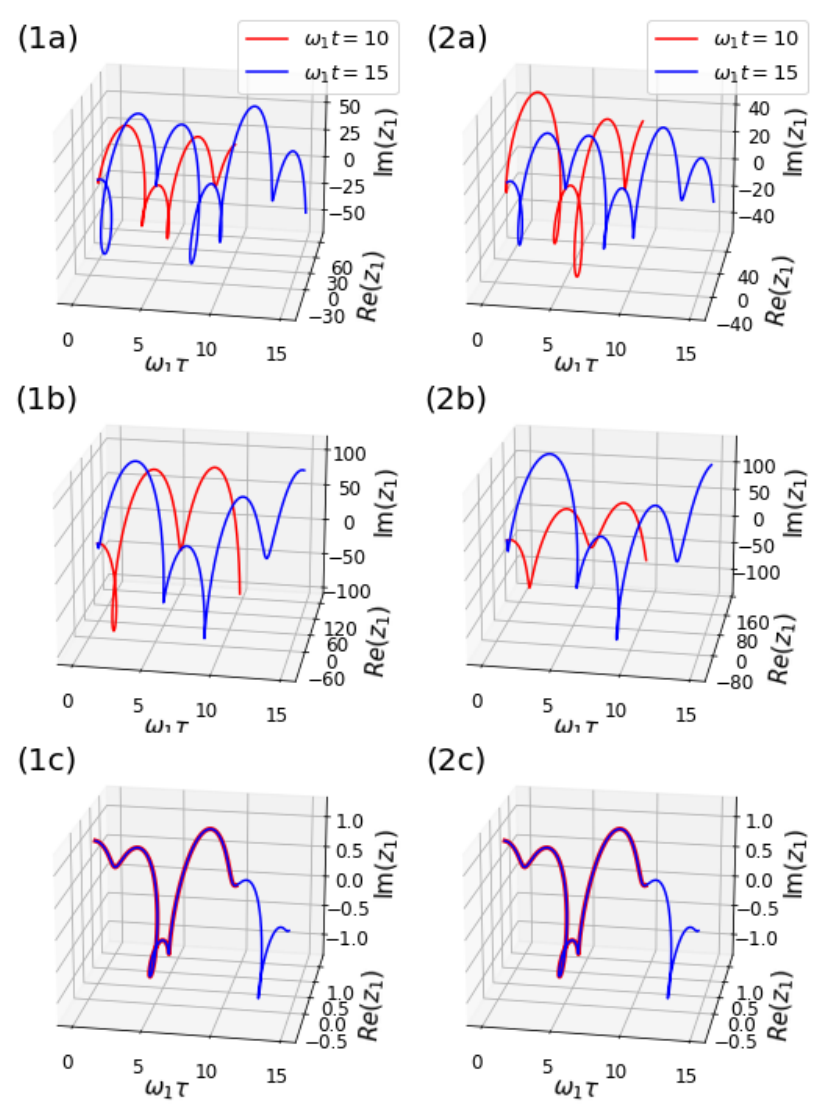}
\caption{(Colour online)  The stationary paths of the dimensionless variable $z_1(\tau)$ for mode 1 in the path integrals of Eq.~(\ref{propagating}), determined  by the equation of motion Eq.~(\ref{eom-z1}) or its solution Eq.~(\ref{sz1}), as a function of time $\tau$ 
varying from the initial time $t_0=0$ to the end (delayed-choice) time $t=10/\omega_1$ or $15/\omega_1$, respectively. 
The left and right panels take arbitrarily two different set of the boundaries (i.e. the initial and final states in the path 
integral) given by $z_{1i}=z'^*_{1i}
=1i; z^*_{1f}=z'_{1f}=2$ and $z_{1i}=1i, z'^*_{1i}=2i ; z^*_{1f}=2, z'_{1f}=1+\sqrt{3}i$, respectively. The plots 
(a1)-(b1) are obtained with $\omega_2=2\omega_1, \gamma=2$, which shows that different later time choices 
produce in advance the different stationary paths, as a direct evidence of causality breaking; the plots (a2)-(b2) 
take $\omega_2=\omega_1, \gamma=2$ which corresponds to the two resonant modes that manifest the similar 
causality breaking effect;
the plots (c1)-(c2) are given by $\omega_2=2 \omega_2$ but $\gamma=0$, namely no squeezing in this case
so that the stationary paths reproduce precisely the classical trajectories for different later time choices, which show 
that the internal causality is preserved. The other parameters  $V_{12}=0.5 \omega_1$ and $\alpha_2=1$. }
\label{fig1}
\end{figure}

However, if $\gamma \neq0$, i.e., the initial state of mode 2 is a coherent squeezed state, which contains some pure quantum effect (squeezing) 
that goes beyond the features of a classical particle, then $\widetilde{g}(\tau,\tau') \neq 0$ and $\overline{g}(\tau,\tau') \neq 0$, see  
Eq.~(\ref{ttcf}). As a result, $v_1(\tau,t) \neq 0$ and $v_2(\tau,t) \neq 0$ such that the causality of the 
stationary paths for each mode is no longer preserved, as shown by Eq.~(\ref{z1 relation}).  
%Here, the pure quantumness of a quantum state can be quantitatively described by Wigner distribution, namely when the Wigner distribution
%of a quantum state occurs negative values, see, for example Ref.~\cite{XZ2015}. It is well-known that a squeezed state has negative values 
%in its Winger distribution. 
Thus, it is the quantumness (here is the squeezing effect) in the initial state of mode 2 induces the influence of the future dynamics ($t>\tau$) 
of mode 1 on its present dynamics at $\tau$ so that the dynamical evolution of each mode is no longer causal. 
In Fig.~\ref{fig1}, we plot a few stationary paths determined by 
Eq.~(\ref{eom-z1}) with two different boundaries (i.e.~two different sets of the fixed starting and ending states) for the 
path integrals of the reduced density matrix operator $\rho_1(t)$. 
The starting states $|z_{1i}\rangle$ for the forward evolution operator can take any state in the whole complex space 
(so does for its dual state  $\langle z'_{1i}|$ for the backward evolution operator). 
The ending states are specified by the reduced density matrix element $\langle z_{1f} |\rho_1(t)|z'_{1f} \rangle$ of Eq.~(\ref{rho_1(t)}) which can be also any arbitrary matrix element of $\rho_1(t)$
in the coherent state representation.  Here, without loss of generality, we take 
the boundaries of (a) $z_{1i}=z'^*_{1i}=1i; z^*_{1f}=z'_{1f}=2$ and (b) $z_{1i}=1i, z'^*_{1i}=2i ; z^*_{1f}=2, z'_{1f}=
1+\sqrt{3}i$, respectively, for the two panels plotted in Fig.~\ref{fig1}. 

As one can see from Fig.~\ref{fig1},  the stationary paths obey the same equation of motion and the same boundaries but 
their trajectories are totally different for different choices of the later time $t$, except for the case with $\gamma=0$ (no squeezing) as presented in Fig.~\ref{fig1}(c1)-(c2). The red and blue paths in each plot correspond to two different later time choices at $t=10/\omega_1$ and $t=15/\omega_1$, respectively. The choice of the time $t$ can be understood as the time point at which one makes an operation on mode 1, such as suddenly turning-off the interaction between the two modes, making a measurement on mode 1 or any other action on mode 1 such that the dynamical evolution of mode 1 is suddenly changed. However, this later action at time $t$ changes the previous stationary path trajectories at an early time $\tau$ ($0< \tau < t$). In other words, different choice of the later time $t$ leads to different past trajectories, as shown by the red and blue paths in Fig.~\ref{fig1}(a1)-(a2)
and (b1)-(b2). This explicitly demonstrates, for the first time in the literature, the breakdown of causality   
in the reduced subsystem dynamics in the quantum mechanical path integral formulation. Only for the situation given by Fig.~\ref{fig1}(c1)-(c2) where 
$\gamma=0$ (no squeezing), the stationary paths follow the same trajectory for different later time choices. It numerically shows that when $\gamma=0$, the quantum evolution gives precisely the same dynamics as the classical one, where the 
causality is preserved, and the corresponding dynamics is identical to the causal classical dynamics, as we have already pointed out. 

The above remarkable results show that the stationary path trajectories of mode 1 are deterministic for both modes being initially in 
Glauber coherent states, but the stationary path trajectories are no longer casual when mode 2 is initially in a quantum states (here is the 
squeezing state, $\gamma\neq 0$). This is because
a different choice of the later time $t$ changes the previous stationary path trajectory, as shown in Fig.~\ref{fig1}(a1)-(a2) and 
(b1)-(b2), due to the internal causality breaking induced by the squeezing effect $\gamma \neq 0$. 
Also, because the values of 
$z_{1i}, z'^*_{1i}$ take over the whole complex space and valid for arbitrary values of $z^*_{1f}, z'_{1f}$, as shown by Eqs.~(\ref{rho_1(t)})-(\ref{propagating}), there are infinite numbers of {\it stationary paths} contributing to the path integrals in Eq.~(\ref{propagating})
when $\gamma\neq 0$. %Notes that this infinite number of the stationary paths is carried out for the first time. 
This naturally 
generates the randomness in the deterministic quantum evolutions. It is worth pointing out that  in Feynman's original path 
integrals the path integral is defined as a sum over all possible paths (an infinity of quantum-mechanically possible paths) 
for the system evolving from one state to another but there is only one stationary path which corresponds to the 
classical trajectory.  When Feynman and Vernon applied the path integral to open quantum systems with linear couplings \cite{Feynman1963}, 
they didn't study further the detailed contributions of the paths in the reduced dynamical evolution, so the above finding was not 
obtained (also in the later investigations by Caldeira and Leggett \cite{Leggett1983}, and others \cite{GSI1988,Weissbook}). 
%{\color{red} In general, “paths” in Feynman’s path integral are usually not physical trajectories that the particle follows, only the entire sum 
%over paths gives the transition amplitudes that makes the measurable predictions. However, in the coherent state path integral formulation,
%no a priori probabilistic interpretation is assumed. The coherent-state path integral functional is an exact expression of the unitary 
%time-evolution operator between two coherent states, derived directly by using the identity resolution of the coherent state. Interesting
%readers can find the detailed derivation given in the textbook by Faddeev and Slavnov \cite{Faddeev1980}. Meantime, we choose the 
%initial state of the system (mode 1) in a coherent state, 
%which is a wave packet with minimum uncertainty, its evolution alone precisely follows the classical trajectories of the harmonic oscillator,
%as we show in Ref.~[\cite{Illustration}]. For linear systems, the sum over paths is fully contributed by the stationary paths.. 
In fact, the causality breaking is the nature of statistics. Thus, the internal causality breaking by the quantumness 
in the deterministic quantum evolution we show above indicates why and how quantum mechanics hold a probability 
interpretation. More will be discussed after we derived exactly the reduced density matrix in the next section (Sec.~IV). 
The connection to the quantum measurement processes is given in Sec.~V.

Also, the solutions given by Eqs.~(\ref{z1 relation}) to (\ref{s_negf}) are valid for the general solutions of open 
quantum systems coupled to an environment containing of infinite number of degrees of freedom.  The only changes 
we need to make is these two-time correlations given in Eq.~(\ref{ttcf}) and the solution of the reduced retarded Green's function given by Eq.~(\ref{uts}). 
More specifically, for a general open quantum system 
described by Eq.~(\ref{Hetot}), i.e.~a simple mode coupled to an environment containing infinite number of modes 
with linear exchange couplings,  these two-time correlations contains the 
contributions from all other modes in the environment as follows,
\begin{subequations}
\label{ttcf_ext}
\begin{align}
g(\tau,\tau') &=  \sum_k\abs{V_{1k}}^2e^{-i\omega_k(\tau-\tau')},  \\
\widetilde{g}(\tau,\tau') &= % \frac{\tanh\gamma^2}{1-\tanh^2\gamma} 
\sum_k \sinh^2\!\gamma_k \abs{V_{1k}}^2e^{-i\omega_k(\tau-\tau')},   \label{sbcc} \\
\overline{g}(\tau,\tau') & = \sum_k \frac{\sinh(2\gamma_k)e^{i(\theta_k+2\omega_kt_0)}}{4} 
% \frac{\tanh \gamma\cdot e^{i(\theta+2\omega_2t_0)}}{1-\tanh^2\gamma}  
 (V_{1k})^2e^{-i\omega_k(\tau+\tau')}.  \label{sbpc}
\end{align}
\end{subequations}
Notice that the solution of the reduced retarded Green's function given by Eq.~(\ref{uts}) is a pure oscillation function oscillating around 
the zero-value, it represents the energy oscillation between the two modes. There is no energy dissipation from the system 
(one mode) to the environment (another mode). When the environment contains infinite number of modes, in particular, continuously distributed modes,
the solution of Eq.~(\ref{uts}) is modified to become a time-dependent damping oscillation function, see the general solution 
given in Refs.~[\cite{Zhang2019,Zhang2012}]. In other words, the reduced retarded Green's function is no longer a pure oscillation function 
oscillating around the zero-value, it becomes a time-dependent damping function describing the non-Markovian dissipation
dynamics in open quantum systems.

On the other hand, Eqs.~(\ref{sbcc}) and (\ref{sbpc}) is assumed that all modes in the environment are initially in squeezed 
coherent states. For the general open quantum systems studied in the literature, usually the environment is initially assumed 
in a thermal equilibrium state (thermal bath) \cite{Feynman1963,Leggett1983,Weissbook}. In this case, Eq.~(\ref{ttcf_ext}) 
is reduced to
\begin{subequations}
\label{ttcf_ext1}
\begin{align}
g(\tau,\tau') &=  \sum_k\abs{V_{1k}}^2e^{-i\omega_k(\tau-\tau')},  \\
\widetilde{g}(\tau,\tau') &= % \frac{\tanh\gamma^2}{1-\tanh^2\gamma} 
\sum_k \frac{1}{e^{\hbar\omega_k/k_BT} +1}\abs{V_{1k}}^2e^{-i\omega_k(\tau-\tau')},   \label{sbcc1} \\
\overline{g}(\tau,\tau') & = 0.  \label{sbpc1}
\end{align}
\end{subequations}
In both cases (the squeezed environment or the thermal bath), the internal causality breaking is manifested in the reduced 
dynamics of the system by its equations of the motion either in terms of the classical trajectories in the complex phase space or 
equivalently in terms of the reduced less Green's function (the reduced correlated Green's function) \cite{Jin_Tu2010}. The reduced 
pairing correlated Green's function is 
need for the squeezing or superconducting environment, as we shown in this work, and also in our previous works \cite{HZ2020,HZ2022b}. 

However,  if one begins with a thermal bath, i.e., let the environment be initially in an equilibrium state which is already built 
on statistical ensembles, then it becomes meaningless to exploring the emergence of statistical probability in quantum mechanics.  
In the previous investigations of open quantum systems with the influence functional \cite{Feynman1963,Leggett1983,HPZ1992,Weissbook}, 
it starts with a thermal bath and therefore lacks the possibility to explore the emergence of probabilistic nature from the beginning.
In the review article \cite{GSI1988}, Grabert, Schramm, and Ingold made an extension of the influence functional to some initially 
correlated (entangled) states. This is also not useful for the motivation proposed in this work, because starting with an initially 
correlated (entangled) states, both the system and the environment are initially in mixed states, it is also meaningless to explore 
the genesis of entanglement and statistical nature from such a construction.

In fact, the nature of internal causality breaking was already presented in our previous works, see the earliest derivation of the 
first fermionic exact master equation \cite{TZ2008}, and then the derivation of exact master equations for various open quantum systems 
later \cite{Jin_Tu2010,LZ2012,LZ2018,HZ2020,HZ2022b}. But we did not make a claim in these early works about the discovery 
of the internal causality breaking and the relation with the emergence of entanglement and statistical probability. 
It is just because in these works, we also assumed that the environment (or environment plus the subsystem) is initially in 
an equilibrium state which is a consequence of statistical ensembles. On the other hand, the mixture of the forward 
and backward propagating paths of the subsystem are crucially relied on the initial states, as we shown in Eqs.~(\ref{ttcf_ext}) and 
(\ref{ttcf_ext1}). For example, if we let the temperature $T=0$ or the squeezed parameter $\gamma_k=0 (s_k=0)$, i.e., the environment 
is initially in a coherent 
state (note that the vacuum state is also a coherent state), then the non-causal terms vanishes immediately in the equations of motion 
given by Eq.~(\ref{eom-1}) for the forward 
and backward propagating paths or Eq.~(\ref{eom-evo}) for the reduced correlated Green's functions. Subsequently the reduced dynamics will 
remain in a causal evolution. 

Thus, starting with a thermal bath or a system-environment entangled state to explore the emergence of entanglement and statistical nature is 
just a circular reasoning. The above analysis should make our original motivation clearer: taking the special setting for the system and the initial states 
is to eliminate the ambiguities and loopholes in addressing the fundamental issue of quantum mechanics. 
%Otherwise, the conclusion will become a paradox on the EPR paradox, rather than a dynamical solution to the EPR paradox. 
Also, we setup the initial state with
one mode in a coherent state and the other in a squeezing coherent state, to precisely put the problem on a clear boundary between the 
classical and quantum physics: without squeezing (let $s=0, i.e.~\gamma=0)$ in Eq.~(\ref{initials}), the system will always evolve in the classical 
world but its states can be described exactly by quantum states (coherent states) and evolves causally. With the squeezing (set $s\neq 0, 
i.e.~\gamma \neq 0)$, we move the dynamics to the quantum world, where the causality of reduced dynamics is broken, causes the emergences 
of both the entanglement and the statistical probability.  This will be demonstrated explicitly in the next section by solving exactly the reduced 
density matrix and its dynamics.

\section{The exact master equation of the reduced density operator and the dynamical genesis of entanglement} \label{sec4}

\subsection{The exact master equation of the reduced density operator}
Now, we are going to show 
that the internal causality breaking shown in the last Section, i.e., all the stationary paths are determined by their past as well as their 
future dynamics, leads to the emergence of entanglement between the two modes, even though the two 
modes are initially unentangled and the coupling of the two modes given 
in Eq.~(\ref{Htot}) alone cannot generates entanglement.  To make this conclusion explicitly, we are now going to solve the reduced density  
operator $\rho_1(t)$ through Eq.~(\ref{rho_1(t)}). 
Using the stationary path equations of motion Eq.~(\ref{eom-1}), we find that the propagating function Eq.~(\ref{propagating}) can be simply reduced to 
\begin{align}
        & \mathcal{J}( z_{1f}  , z'_{1f},t; z_{1i},z'_{1i},t_0) \!= \notag \\
        &~ \Bar{N}\!(t)\! \exp \!\bigg\{\!\frac{1}{2}\big[{z'_1}^\ast\!(t)z'_{1f}\!+\!z^\ast_{1f}z_1(t)\!+\!z^\ast_1(t_0)z_{1i}\!+\!{z'}^\ast_{1i}z'_1(t_0)  \notag \\
        &~~ +\!(z^\ast_{1f}\!-\!{z'_1}^\ast\!(t))v_0(t,t_0)\alpha_2\!-\! \alpha^\ast_2 v_0^\ast(t,t_0)(z_1(t)\!-\!z'_{1f})\big]\!\bigg\} ,
        \label{propagating'}
\end{align}
where $\bar{N}(t)$ is related to the normalized factor determined by the condition $\Tr_{1+2}[\rho_{tot}(t)]=\Tr_1[\rho_1(t)]=1$, as given
explicitly in the next equation.
The end-point values of the stationary paths $z_1(t)$, $z^\ast_1(t_0)$, $z'_1(t_0)$ and ${z'_1}^\ast(t)$ are determined from Eq.~(\ref{z1 relation}).

Using the relations of Eq.~(\ref{z1 relation}), the above solution of the propagating function can be further expressed as
%\begin{widetext}
\begin{align}
        \mathcal{J}(& z_{1f} , z'_{1f},t; z_{1i},z'_{1i},t_0) \!  \notag \\
        & =\!\Bar{N}(t)\exp \!\bigg\{ \!\!
             \begin{pmatrix}
            z^\ast_{1f} & z'_{1f}
        \end{pmatrix} \!\!
        \Big[\bm {J}_1(t)
       % \begin{pmatrix}
        %    \frac{u(t,t_0)(1+v_1(t,t))}{\Delta(t,t)} & \frac{u^\ast(t,t_0)v_2(t,t)}{\Delta(t,t)}\\
        %    \frac{u(t,t_0)v^\ast_2(t,t)}{\Delta(t,t)} & \frac{u^\ast(t.t_0)(1+v_1(t,t))}{\Delta(t,t)}
        %\end{pmatrix}\!\!
       \! \begin{pmatrix}
            z_{1i} \\  z'^*_{1i}
        \end{pmatrix}  
       \! + \!  \bm{J}_2(t) \!
         \begin{pmatrix}
            \alpha_2 \\  \alpha_2^\ast
        \end{pmatrix} \! \Big] \notag \\
        & \qquad+ \! \Big[ \!
         \begin{pmatrix}
            \alpha^\ast_2 & \alpha_2
        \end{pmatrix}\! 
        \bm{J}_1(t)  \!+\!  \frac{1}{2}\! \begin{pmatrix}
            z'^*_{1i} & z_{1i}
        \end{pmatrix}\!
        \bm{J}_3(t) \Big]\!
       % \begin{pmatrix}
        %    1\!-\!\frac{\abs{u(t,t_0)}^2(1+v_1(t,t))}{\Delta(t,t)} & -\frac{(u^\ast(t,t_0))^2v_2(t,t)}{\Delta(t,t)}\\
        %    -\frac{u^2(t,t_0)v^\ast_2(t,t)}{\Delta(t,t)} & 1\!-\!\frac{\abs{u(t,t_0)}^2(1+v_1(t,t))}{\Delta(t,t)}
       % \end{pmatrix}\!\!
       \!  \begin{pmatrix}
            z_{1i} \\  z'^*_{1i}
        \end{pmatrix}\notag\\
        & \qquad \qquad \quad +\!\frac{1}{2}\!\begin{pmatrix}
            z^\ast_{1f} & z'_{1f}
        \end{pmatrix}\!\!
        \bm{J}_4(t) \!
       % \begin{pmatrix}
        %    1\!-\!\frac{1+v_1(t,t)}{\Delta(t,t)} & -\frac{v_2(t,t)}{\Delta(t,t)}\\
        %    -\frac{v^\ast_2(t,t)}{\Delta(t,t)} & 1\!-\!\frac{1+v_1(t,t)}{\Delta(t,t)}
       % \end{pmatrix}\!\!
        \begin{pmatrix}
            z'_{1f} \\  z^\ast_{1f}
        \end{pmatrix}
     % \notag \\
 %       &+\!
 %      \begin{pmatrix}
 %           \alpha^\ast_2 & \alpha_2
 %       \end{pmatrix}\!\!
 %      \bm{J}_4(t)
        %\begin{pmatrix}
        %     \frac{v^\ast_0(t,t_0)(1+v_1(t,t))}{\Delta(t,t)} &\frac{v^\ast_0(t,t_0)v_2(t,t)}{\Delta(t,t)}\\
        %    \frac{v_0(t,t_0)v^\ast_2(t,t)}{\Delta(t,t)} & \frac{v_0(t,t_0)(1+v_1(t,t))}{\Delta(t,t)}
        %\end{pmatrix}\!\!
 %      \begin{pmatrix}
 %           z'_{1f} \\  z^\ast_{1f}
 %       \end{pmatrix}\!-\!
 %       \begin{pmatrix}
 %           \alpha^\ast_2 & \alpha_2
 %       \end{pmatrix}\!\!
 %       \bm{J}_5(t)
        %\begin{pmatrix}
        %    \frac{u(t,t_0)(1+v_1(t,t))}{\Delta(t,t)} & \frac{u^\ast(t,t_0)v_2(t,t)}{\Delta(t,t)}\\
        %    \frac{u(t,t_0)v^\ast_2(t,t)}{\Delta(t,t)} & \frac{u^\ast(t.t_0)(1+v_1(t,t))}{\Delta(t,t)}
        %\end{pmatrix}\!\!
 %       \begin{pmatrix}
 %            z_{1i} \\  {z'_{1i}}^\ast
 %      \end{pmatrix}\!
 \!\! \bigg\}
\end{align}
%\end{widetext}
where 
\begin{subequations}
\begin{align}
\bm {J}_1(t) = &
\begin{pmatrix}
          w_1(t)u(t,t_0) & w_2(t) u^\ast(t,t_0)\\
            w^*_2(t)u(t,t_0)& w_1(t) u^\ast(t.t_0)
        \end{pmatrix} ,  \\
\bm {J}_2(t) = &
        \begin{pmatrix}
             w_1(t)v_0(t,t_0) &w_2(t)v^\ast_0(t,t_0)\\
            w^*_2(t)v_0(t,t_0)& w_1(t)v^*_0(t,t_0)
        \end{pmatrix} ,  \\
\bm {J}_3(t) = &
        \begin{pmatrix}
            1\!-\!w_1(t)\abs{u(t,t_0)}^2 & -w_2(t)(u^\ast(t,t_0))^2\\
            -w^*_2(t)u^2(t,t_0) & 1\!-\!w_1(t)\abs{u(t,t_0)}^2
        \end{pmatrix} , \\
\bm {J}_4(t) = &
         \begin{pmatrix}
            1\!-\!w_1(t) & -w_2(t)\\
            -w^*_2(t) & 1\!-\!w_1(t)
        \end{pmatrix} 
%&\bm {J}_5(t)=  
%        \begin{pmatrix}
%            \frac{u(t,t_0)(1+v_1(t,t))}{\Delta(t,t)} & \frac{u^\ast(t,t_0)v_2(t,t)}{\Delta(t,t)}\\
%            \frac{u(t,t_0)v^\ast_2(t,t)}{\Delta(t,t)} & \frac{u^\ast(t.t_0)(1+v_1(t,t))}{\Delta(t,t)}
%        \end{pmatrix}\!\!
\end{align}
\end{subequations}
with $w_1(t)=[1+v_1(t,t)]/\Delta_2(t)$, $w_2(t)=v_2(t,t)/\Delta_2(t)$ and $\Delta_2(t)=(1+v_1(t,t))^2-\abs{v_2(t,t)}^2$. 
The solutions of the reduced Green's functions $u(t,t_0),v_0(t,t_0), v_1(t,t)$ and $v_2(t,t)$ are given by Eq.~(\ref{s_negf}).

Then, utilizing the same approach we developed for deriving the exact master equation of open quantum systems \cite{LZ2012,Zhang2019}, 
we obtain the exact 
equation of motion for the reduced density matrix $\rho_1(t)$ for this two mode coupled system as the simplest open quantum system,
\begin{align}
\frac{d}{dt} & \rho_1 (t)   =  \frac{1}{i\hbar} [H'_s(t), \rho_1(t) ] \notag \\
&+ \! \lambda (t)\big\{a_1\rho_1(t) a^\dag_1 \!-\! \frac{1}{2}a^\dag_1 a_1\rho_1(t) \!-\! \frac{1}{2}\rho_1(t)a^\dag_1 a_1\big\}\notag\\
& + \! \widetilde{\lambda}(t)\big\{a_1\rho_1(t) a^\dag_1 \!+\! a^\dag_1 \rho_1(t) a_1 \!-\! a^\dag_1 a\rho_1(t) \!-\! \rho_1(t)a_1 a^\dag_1 \big\} \notag \\
& + \! \overline{\lambda}(t) \big\{ a^\dag_1 \rho_1(t) a^\dag_1 \!-\! \frac{1}{2} a^\dag_1 a^\dag_1 \rho_1(t) \!-\! \frac{1}{2} \rho_1(t) a^\dag_1 a^\dag_1  \big\} \notag \\
& + \! \overline{\lambda}^*(t) \big\{ a_1 \rho_1(t) a_1 \!-\! \frac{1}{2}a_1 a_1 \rho_1(t) \!-\! \frac{1}{2} \rho_1(t) a_1 a_1  \big\} .
\label{merho1}
\end{align}
which is valid for arbitrary initial state $\rho_1(t_0)$ of the first mode. Here 
\begin{subequations}
\begin{align}
H'_s(t)= \hbar\omega'_1(t) a^\dag_1 a_1 + \hbar f_1(t) a^\dag_1 + \hbar f^*_1(t)a_1
\end{align}
is the renormalized Hamiltonian of mode 1 with the renormalized frequency and an effective driving field as follows 
\begin{align}
& \omega'_1(t)= -  \text{Im}\bigg[\frac{\dot{u}(t,t_0)}{u(t,t_0)} \bigg] , \\
& f_1(t)=i\alpha_2 \Big[ \dot{v}_0(t,t_0)-\frac{\dot{u}(t,t_0)}{u(t,t_0)}v_0(t,t_0)\Big] .
\end{align}
\end{subequations}
These effects are induced by the coupling between the two modes and the coherent part of the initial state of mode 2 after taken the trace over 
all the states of the mode 2.  The coefficients $\lambda(t)$, $\widetilde{\lambda}(t)$ and $\overline{\lambda}(t)$ in Eq.~(\ref{merho1}) 
are given by the following relations,
\begin{subequations}
\label{mepara}
\begin{align}
\lambda(t) & =- 2 \text{Re}\!\left[\frac{\dot{u}(t,t_0)}{u(t,t_0)}\right]  \notag \\
&=\frac{1}{2}(\omega_{+}\!-\!\omega_{-})\sin^2\!\varphi \sin\left[(\omega_{+}\!-\!\omega_{-})(t-t_0)\right] ,  \label{dissip} \\
\widetilde{\lambda}(t) & =\dot{v}_1(t,t)-2v_1(t,t)\text{Re}\!\left[\frac{\dot{u}(t,t_0)}{u(t,t_0)}\right] ,  \label{flus1} \\
\overline{\lambda}(t) &=\dot{v}_2(t,t)-2\frac{\dot{u}(t,t_0)}{u(t,t_0)}v_2(t,t) . \label{flus2}
\end{align}
\end{subequations}
%which are induced by the coupling between the two modes as well as the squeezing part of the initial state of mode 2 
%after taken the trace over all the states of the mode 2. If there is no coupling between the two modes, i.e.~$V_{12}=0$, then 

Equation (\ref{merho1}) shows that the equation of motion (master equation) for the reduced density operator $\rho_1(t)$ consists of  two parts, 
a unitary part plus a 
non-unitary part, which is a general structure of the master equation for open quantum systems, %\cite{Leggett1983,HPZ1992,Weissbook,TZ2008,LZ2012,Zhang2019,HZ2022b},
and it is very different from the unitary Schr\"{o}dinger evolution of Eq.~(\ref{Sche}) 
or the von Neumann evolution of Eq.~(\ref{vNe}) for the coupled two mode system as an closed system. 
The unitary part in Eq.~(\ref{merho1}) is govern by the renormalized Hamiltonian $H'_s(t)$, while the non-unitary part is induced by 
the coupling between the two modes and the squeezing effect in the initial state of mode 2. If there is no coupling between the two 
modes, i.e.~$V_{12}=0$, then all the coefficients vanishes, i.e.~$f_1(t)=\lambda(t)=\widetilde{\lambda}(t)=\overline{\lambda}(t)=0$ 
and $\omega'_1=\omega_1$ (which can be seen clear that the reduced correlated Green's functions $v_0(t,t)=v_1(t,t)=v_2(t,t)=0$ when $V_{12}=0$). 
Thus, Eq.~(\ref{merho1}) is naturally reduced to a unitary equation. 
On the other hand, if the squeezing parameter  in the initial state of mode 2 vanishes, namely $\gamma=0$ ($s=0$), even if the two modes 
have the coupling $V_{12}\neq 0$, we also have $v_1(t,t)=v_2(t,t)=0$ so that  
$\widetilde{\lambda}(t)=\overline{\lambda}(t)=0$ but $f_1(t) \neq0$ and $\lambda(t) \neq 0$. 
%In this case, Eq.~(\ref{merho1}) is still a non-unitary equation of motion. This is because Eq.~(\ref{merho1}) is valid for arbitrary i
%nitial state $\rho_1(t_0)$. Only 
When both modes are in Glauber coherent states, $\rho_1(t)$ and so does $\rho_2(t)$ remains in a pure state and maintains the unitary
evoluytion.  We will show explicitly this conclusion next.

In any case, the total density matrix operator $\rho_{\rm tot}(t)$ always keeps in a pure state for all the time if its initial state is a pure 
state. This is because $\rho_{\rm tot}(t)$ is determined by the unitary evolution equation (\ref{ted}) or equivalently the 
Schr\"{o}dinger equation for all the time.
In fact, 
%Eq.~(\ref{merho1}) for the reduced density matrix $\rho_1(t)$ has the same form as the exact master equation 
%for various open quantum systems (including bosonic, fermionic and topological open systems) we developed in the last two 
%decades \cite{TZ2008,LZ2012,LZ2018,Zhang2019,HZ2022b}, using the approach based on
%Feynman-Vernon's influence functional \cite{Feynman1963}. 
the breakdown of the unitary for the time evolution of the reduced density matrix operator $\rho_1(t)$ is already manifested 
in the effective action for mode 1 in the influence functional, as shown by Eqs.~(\ref{IFs}) and (\ref{IFes}) where the effective action 
contains a real part and an imaginary part. It is the imaginary part in the effective 
action that results in the breakdown of the unitary. This conclusion was indeed first obtained by Feynman and Vernon \cite{Feynman1963} 
and Caldeira and Leggett \cite{Leggett1983} in the study of quantum Brownian motion \cite{Weissbook}, also see our general studies of 
open quantum systems \cite{TZ2008,LZ2012,LZ2018,Zhang2019,HZ2022b}. 
But it has never been recognized that even for a simple system with two coupled modes, the 
quantum evolution of a subsystem (one of the two modes) contains the complicated dynamics of a much more complicated 
open system. 

It is worth mentioning that Eq.~(\ref{merho1}) has some different dynamics from the usual open 
quantum system master equation we obtained before \cite{TZ2008,LZ2012,LZ2018,Zhang2019,HZ2022b}.
The most significant difference is the coefficient $\lambda(t)$ given by Eq.~(\ref{dissip}), which is a pure oscillating function around  
the zero point. This indicates that there is not dissipation for this simple coupled two mode system (or any composite 
systems with finite degrees of freedom). The oscillation represents energy and information exchange back and forward 
between two modes without energy decay from one mode to the other, representing the information exchange between 
different modes in all the time 
as shown by Eq.~(\ref{dissip}). Only when the system couples to an environment with continuous spectra, 
can dissipation or damping occur \cite{Zhang2012}. 

On the other hand, the coefficients $\widetilde{\lambda}(t)$ and $\overline{\lambda}(t)$ are induced by the initial squeezing of the mode 2, 
rather than the thermal fluctuations given in the usual studies of open quantum systems.  It is these terms that breaks the unitary 
of the evolution for the reduced density matrix $\rho_1(t)$, even if the thermal fluctuations absence.
%, which is obvious because they are superoperators with Lindblad forms \cite{LZ2012,HZ2022b}. 
%In fact, if we let the mode 1 coupled to continuous modes, namely, replace the Hamiltonian of Eq.~(\ref{Htot}) by 
%$H_{tot}=\hbar\omega_{1}a_{1}^\dag a_{1}\!+\!\sum_k \hbar\omega_{k} a_{k}^\dag a_{k}\!+\!\sum_k \hbar(V_{1k}a_{1}^\dag 
%a_{k}\!+\!V^\ast_{1k}a_{k}^\dag a_{1})$, then we only need to make a corresponding change of $V_12$ to $V_1k$ 
%in Eq.~(\ref{} and sum over all the continuous k-mode, the equation of motion for the reduced density matrix ....
The above analysis show that the general theory of open quantum systems can be applied to the very simple system 
consisting of only two modes through a simple exchange interaction, 
and the non-unitarity can emerge in the subsystem evolution in such simple systems, as long as the initial states contains pure 
quantumness (squeezing in the present case).
%Also note that quantum fluctuations in quantum mechanics exist intrinsically due to the non-commutativity of two 
%corresponding operators, as von Neumann realized in 1926.  
This non-unitary behavior can be generalized to the evolution of individual particle (subsystems) in any interacting composite system, 
even though the whole system is isolated and obeys the dynamical unitary evolution of the Schr\"{o}dinger equation Eq.~(\ref{Sche}) 
or the von Neumann equation Eq.~(\ref{vNe}).
%This conclusion seems to have not been widely recognized in the literature, in particular in the current development of 
%quantum technology. It indicates that manipulation of  true unitary operations for individual qubits in a coupled many-qubit 
%systems is in principle extremely difficulty to be implemented, even if the decoherence . 

\subsection{Dynamical genesis of entanglement and the emergence of statistical probability}
Now, we consider the initial state  $\rho_1(t_0)=|\alpha_1\rangle \langle \alpha_1|$, i.e. the system is initially in a classical state that
can be precisely described by a quantum state \cite{ZhangRMP1990}, as 
given in Eq.~(\ref{initials}). It is straightforward to find the analytical reduced density matrix operator $\rho_1(t)$ in the 
coherent state representation 
%{\color{red}
\begin{align}\label{rho_1(t)m}
%\begin{split}
    \langle z_{1f}|\rho_1(t)|  z'_{1f}  \rangle  &\!=\!  N(t)\exp\Big\{\delta(t)[z_{f}^\ast\!-\!\alpha_1^\ast(t)][z'_{f}\!-\!\alpha_1(t)]\Big\}   \notag \\
    &\!\times\! \exp\Big\{z_{1f}^\ast\alpha_1(t)\!-\!\frac{1}{2}\beta(t)[z_{1f}^\ast\!-\!\alpha_1^\ast(t)]^2\Big\}   \notag \\
    &\!\times\! \exp\Big\{\alpha_1^\ast(t) z'_{1f}\!-\!\frac{1}{2}\beta^*(t)[z'_{1f}\!-\!\alpha_1(t)]^2\Big\} 
%\end{split}
\end{align}
where $\alpha_1(t)=u(t,t_0)\alpha_1+v_0(t,t_0)\alpha_2$, and
\begin{align}
\label{coe_rho1}
\begin{split}
    &\beta(t)=\frac{\tanh\gamma e^{i\theta}v_0^2(t,t_0)}{1-(1-\abs{v_0(t,t_0)}^2)^2\tanh^2\gamma}\\
    &\delta(t)=\frac{\tanh^2\gamma(1-\abs{v_0(t,t_0)}^2)\abs{v_0(t,t_0)}^2}{1-(1-\abs{v_0(t,t_0)}^2)^2\tanh^2\gamma} \\
    &N(t)=\frac{\sech\gamma \exp(-\abs{\alpha_1(t)}^2)}{\sqrt{1-(1-\abs{v_0(t,t_0)}^2)^2\tanh^2\gamma}} .
\end{split}
\end{align}
In Eq.~(\ref{rho_1(t)m}), the term $\delta(t)[z_{2f}^\ast-\beta^\ast(t)][z'_{2f}-\beta(t)]$  in the middle exponent factor or more precisely, 
the term $\delta(t)z_{2f}^\ast z'_{2f}$ in the exponent makes the reduced density matrix operator impossible to be written as the external 
product of a pure state, as a consequence of the non-unitarity of Eq.~(\ref{merho1}). In other words, $\rho_1(t)$ must be a mixed state 
if the coefficient $\delta(t)$ is not zero. On the other hand,
the total density matrix operator of the two modes is guaranteed to remain in a pure state under the unitary quantum evolution of Eq.~(\ref{unitary_evolution}) 
governed by Eq.~(\ref{Htot}). This is  because the initial state of Eq.~(\ref{initials}) is a pure state and the total system is isolated. 
Thus, it provides a direct proof how the total state of this coupled two-mode system becomes a pure entanglement state during the 
dynamical evolution. When one mode is in a mixed state, the other one must be a mixed state, because the total state remains in a pure
state. Thereby, the total density matrix operator at later time $t$ cannot be written as a direct product of the two 
individual mode states, $\rho_{tot}(t) \neq \rho_1(t) \otimes \rho_2(t)$ if $\gamma\neq 0$. 

More explicitly, we can write down the exact reduced density matrix operator $\rho_1(t)$ from Eq.~(\ref{rho_1(t)m}) without 
relying on the coherent state representation,
\begin{align}\label{rho_1(t) sol}
   \rho_1&(t)  =\tilde{N}(t) e^{A^\dag (t)}
    \bigg[\sum^\infty_{n=0} \delta^n(t)|n\rangle \langle n|\bigg] e^{A(t)}
\end{align}
with $\tilde{N}(t)\!=\! N(t) \exp\big\{\delta(t)\abs{\alpha_1(t)}^2 \!-\! \frac{1}{2}[\beta^*(t)\alpha_1(t)^2\! + {\rm h.c.}]\big\}$, 
$A^\dag(t) \!\equiv \! [(1\!-\!\delta(t))\alpha_1(t) \!+\! \beta(t)\alpha_1^*(t)]a^\dag_1\!-\!\frac{1}{2}\beta(t)a^{\dag 2}_1$,
and $|n\rangle= \frac{1}{\sqrt{n!}} (a^\dag_1)^n|0\rangle$ is the Fock state.
The summation $\sum^\infty_{n=0}\delta^n(t)|n\rangle\langle n|$ is indeed a thermal-like state when $\delta(t)$ is not zero.
Thus, the state $\rho_1(t)$ is obviously a mixed state. We can also re-express the reduced density matrix $\rho_1(t)$ as 
\begin{align}\label{simplifed}
    \rho_1(t)\!=\!\mathcal{N}(t)\!\sum^\infty_{n=0}\!\!\frac{\delta^n(t)}{n!}(a^\dag_1)^n|\tilde{\alpha}(t),\xi(t)\rangle\langle \tilde{\alpha}(t),\xi(t)|(a_1)^n\!,
\end{align}
where $|\tilde{\alpha}(t),\xi (t)\rangle$ is a squeezed coherent state. The coherent parameter $\tilde{\alpha}(t)=\frac{1}{1-\abs{\beta(t)}^2}
(c^\ast(t)-\beta(t) c(t))$ with $c(t)\!=\![1-\delta(t)]\alpha_1^\ast(t)+\beta^*(t)\alpha_1(t)$, and the squeezing parameter 
$\xi(t)=\bar{\gamma}e^{i\bar{\theta}}$ with $\bar{\gamma}=\frac{1}{2}\ln\frac{1-\abs{\beta(t)}}{1+\abs{\beta(t)}}$ 
and $\bar{\theta}\!=\!\arg(-\frac{\beta(t)^\ast}{|\beta|})$. In fact, Eqs.~(\ref{rho_1(t) sol}) 
and (\ref{simplifed}) are usually called as squeezed "thermal" state or "thermalized" squeezing state. Here "thermal" or "thermalized"
only means the mixture of the states, unless we extend the system to couple to an infinite number of other modes \cite{HZ2022}. 

The reason that the reduced density matrix operator $\rho_1(t)$ of mode 1 becomes a mixed state (i.e.~the two modes are 
entangled) through their dynamical evolution is because the initial state of mode 2 contains quantumness so that the 
causality of the dynamical evolution of each mode is {\it internally} broken through the forward and backward propagating  
path mixing, as shown by Eq.~(\ref{eom-1}) and Fig.~\ref{fig1}.  If mode 2 is  initially also in a coherent state as mode 1, 
namely both modes are initially in coherent states which correspond to minimum-uncertainty  wave packets for well-defined 
classical particles, the reduced density matrix $\rho_1(t)$ 
will keep to stay in pure states (more precisely, coherent states), and the entanglement 
 between the two modes never emerge. This can be easily justified by setting the squeezed parameter $s=0$ 
 (i.e., $\gamma=0$) in the initial state of Eq.~(\ref{initials}). This setting immediately leads to the coefficients $\beta(t)=0$ 
 and $\delta(t)=0$ in Eq.~(\ref{coe_rho1}). Thus, the reduced density matrix $\rho_1(t)$  in Eq.~(\ref{rho_1(t)m}), (\ref{rho_1(t) sol}) 
 and (\ref{simplifed}) is simply reduced to 
\begin{align}
    \rho_1(t)=D(\alpha_1(t))|0\rangle\langle 0|D^\dag(\alpha_1(t))=|\alpha_1(t)\rangle\langle\alpha_1(t) |,  \label{rdmps}
\end{align} 
where $\alpha_1(t) = u(t,t_0)\alpha_1 + v_0(t,t_0)\alpha_2$.
This is a pure state and evolves exactly as a classical harmonic oscillator, no entanglement emerges. It describes precisely 
the same evolution of a classical harmonic oscillator coupled with another oscillator. The trajectory is given by $\alpha(t)$ in the complex phase space,
in which the causality is preserved. This agrees with the stationary path solution of Eq.~(\ref{clas}).  It also answers the question why 
classical deterministic dynamics obey the causality and classical physics cannot emerge entanglement. 

%The above results further indicates that the initial quantumness induces the internal causality breaking in composite systems.
%It also results in the emergence of entanglement between the subsystems. In fact, any realistic system cannot be isolated 
%so that any measured system must interacting  .....

\begin{figure}
\includegraphics[width=0.9 \linewidth]{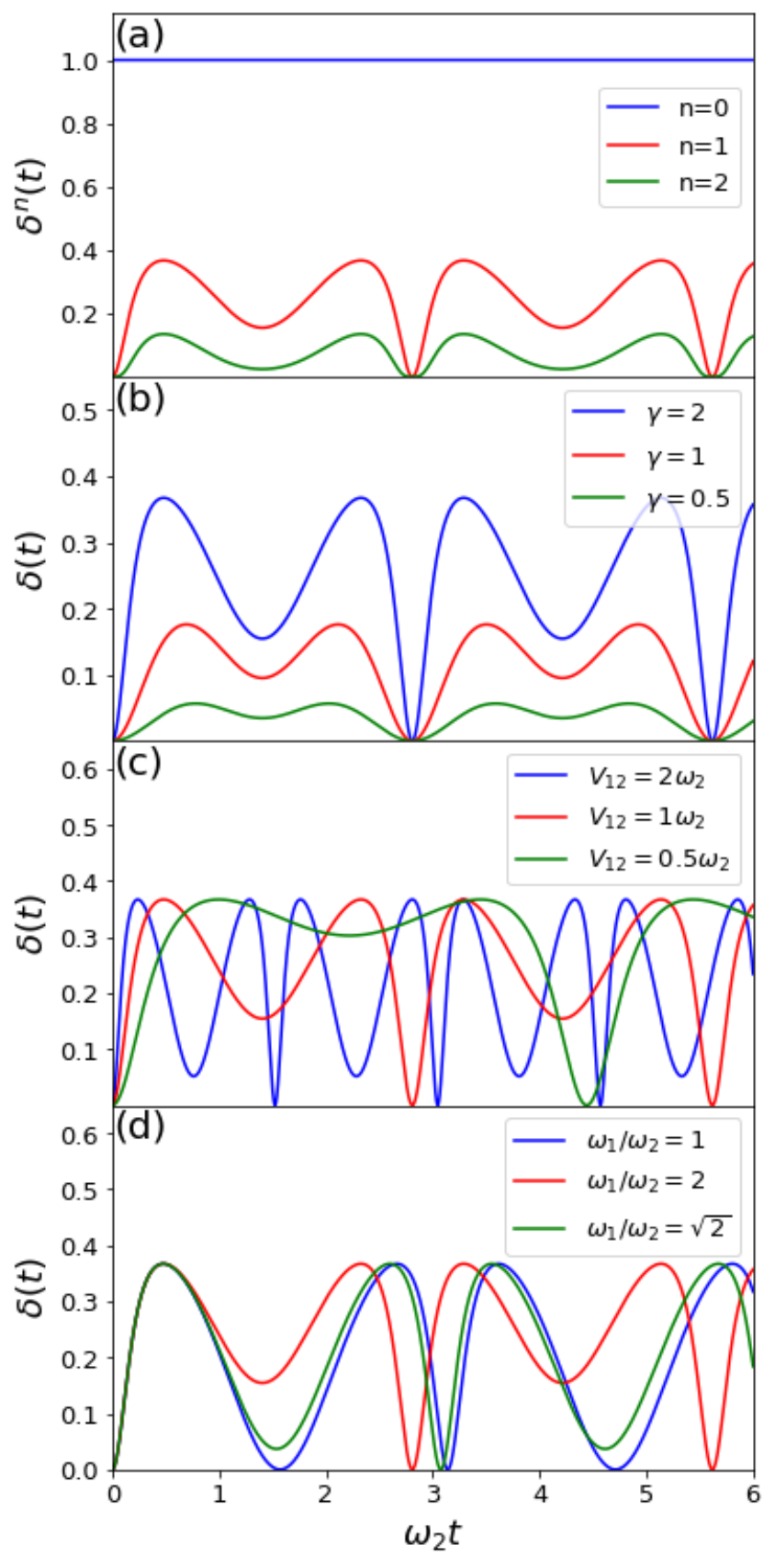}
\caption{(Colour online)  The dimensionless coefficient $\delta(t)$ in Eq.~(\ref{rho_1(t)m})-(\ref{simplifed}) as a function of time, 
which characterizes the mixing degree of the reduced density matrix $\rho_1(t)$. 
(a) The first few terms in the expansion of Eq.~(\ref{rho_1(t) sol}), $\delta^n(t)$ with $n=0,1, 2$, the other parameters in the calculation are 
$(\omega_1/\omega_2)=2$, $V_{12}= \omega_2$, and the squeezed parameter $\gamma=2$. (b-d) $\delta(t)$ for different 
squeezed parameter $\gamma=2, 1, 0.5$,  the different coupling strength parameters  $V_{12}=(2, 1, 0.5) \omega_2$, 
and the different frequencies setting $(\omega_1,\omega_2)= (1,1), (2, 1), (\sqrt{2},1)$, respectively.}
\label{fig2}
\end{figure} 

To see how the reduced density matrix $\rho_1(t)$ changes in time as a mixed state, which is
characterized by $\delta(t) \neq 0$ when $\gamma \neq 0$, we present some numerical results 
in Fig.~\ref{fig2} for the values of $\delta(t)$ in time for various different cases. In Fig.~\ref{fig2}(a)
we show the first few values of $\delta^n(t)$ in the expansion of Eq.~(\ref{rho_1(t) sol}) [or Eq.~(\ref{simplifed})]. 
It shows that the mixed state of mode 1 (also representing the entanglement between the two modes)
change periodically in time. It becomes clearer in Fig.~\ref{fig2}(b-d), where $\delta(t)$ is plotted as a 
function of time for the different squeezed parameters, different coupling strengths, and different set
of the two mode frequencies, respectively. It always shows the periodicity, which is the essence for finite 
quantum systems. As long as the system contains finite number of particles or modes, the
evolution is always periodic.  The squeezing strength $\gamma$ affects the oscillation amplitude, as shown in 
Fig.~\ref{fig2}(b). The two-mode coupling strength $V_{12}$ changes the oscillation frequencies, as given 
by Fig.~\ref{fig2}(c), which corresponding to the two normal mode frequencies in Eq.~(\ref{uts}). Because of
the periodicity, there is still a discrete set of points with $\delta(t) = 0$ where the states of both modes
retrieve to be a pure state and no entanglement at these points, but the measure of the set is zero. 
In the reality, many other modes exist in the surrounding,  different frequencies will generate different normal 
mode frequencies which causes different periodicities, as shown in Fig.~\ref{fig2}(d). Thus, when one oscillator 
couples to more and more other modes, the periodicities of $\delta(t)$ will be eventually merged and disappear. This is 
in fact a general procedure of how a subsystem in a complicated composite system is eventually 
thermalized \cite{HZ2022}, even though the initial system of the total system is a pure state here. The underlying feature 
behind the thermalization is indeed the internal causality breaking for every constitutes or subsystems in a large 
composite system. This is the foundation of statistical mechanics and thermodynamics. %We will leave this problem in a further investigation.  

The above solutions, Eq.~(\ref{rho_1(t)m}) to Eq.~(\ref{rdmps}), shows that at a simple condition (no squeezing or no pure quantumness), 
the quantum dynamics precisely reproduces the classical dynamics, and under certain circumstances (where squeezing or any other 
pure quantumness occur) the entanglement emerges through the quantum dynamical evolution with internal causality breaking.
We should also point out that the motivation to set up $\alpha_1,  \alpha_2  \neq 0$ in this work is to address the basic issue 
arisen from the EPR paradox, i.e., why entanglement emerges only in quantum realm but not in classical world. 
In the seminal EPR paper \cite{einstein1935can}, Einstein {\it et al.}~used a plane wave wavefunction which is quantum 
mechanically unphysical \cite{Tannoudji1991} so that the differences between quantum and classical dynamics cannot be manifested 
and distinguished.  

Here we can see that when we setup the initial state with one mode in a coherent state and the other in a squeezing coherent state, 
the system is precisely putted  on a clear boundary between the classical and quantum physics. The ambiguities in the EPR paradox
are removed. Thus,  by changing the squeezing
parameter of mode 2 from a zero value to a non-zero value, we demonstrate how the dynamics of mode 
1 changes from the causal classical evolution into the non-causal quantum evolution, where the non-causal evolution is induced
by the mixture of the forward and backward propagating paths as shown by Eq.~(\ref{eom-1}). Physically, it is manifested in the equation of
motion for the reduced correlated Green's function $v_1(\tau, t)$ and $v_2(\tau,t)$ given in Eq.~(\ref{eom-evo}). These two reduced Green's functions
$v_1(\tau,t)=\langle a^\dag_1(t) a_1(\tau) \rangle - \langle a^\dag_1(t)\rangle \langle a_1(\tau) \rangle$ and $v_2=\langle a_1(t) a_1(\tau)\rangle
-\langle a_1(t) \rangle\langle a_1(\tau) \rangle$, 
are typically the coherent functions of mode 1 in quantum optics that are experimentally measurable. 

Furthermore, even if mode 1 and mode 2 are initially in a vacuum state and 
a vacuum squeezing state, respectively (i.e., let $\alpha_1= \alpha_2 =0$), the two modes will still soon entangle together.  
Explicitly, let $\alpha_1=\alpha_2 =0$, then $\alpha(t)=0$ in Eq.~(\ref{rho_1(t)m}). Thus, Eqs.~(\ref{rho_1(t) sol}) and 
(\ref{simplifed}) are simply reduced to 
\begin{align}
\label{rho_1(t) sol_c0}
   \rho_1(t) &=N_0(t) e^{-\frac{1}{2}\beta(t)a_1^{\dag 2}}\bigg[\sum^\infty_{n=0} \delta^n(t)|n\rangle \langle n|\bigg] 
   e^{-\frac{1}{2}\beta^\dag (t)a^2_1} \notag \\
   &=\mathcal{N}_0(t)\sum^\infty_{n=0}\frac{\delta^n(t)}{n!}(a^\dag_1)^n| \xi(t)\rangle\langle \xi(t)|(a_1)^n
\end{align}
where $N_0(t)=\sech\gamma/\sqrt{1-(1-\abs{v_0(t,t_0)}^2)^2\tanh^2\gamma}$. This is still a mixed state so that
the coupled two modes are entangled through the dynamical evolution.  
Experimentally, starting with a squeezing 
light source and a beam splitter, measuring the quadratures, one can observe the properties of the above mixed state and 
then demonstrate the genesis of entanglement as well as the emergence of statistical probability in quantum mechanics. 
%But it should be pointed out that this does not mean a vacuum mode is entangled with a squeezing mode 
%because the initial vacuum state  no longer remains in the vacuum under its dynamical evolution through the 
%beam splitting with another mode, as shown in Eq.~(\ref{rho_1(t) sol_c0}).

\begin{figure}
\includegraphics[width=1.0\linewidth]{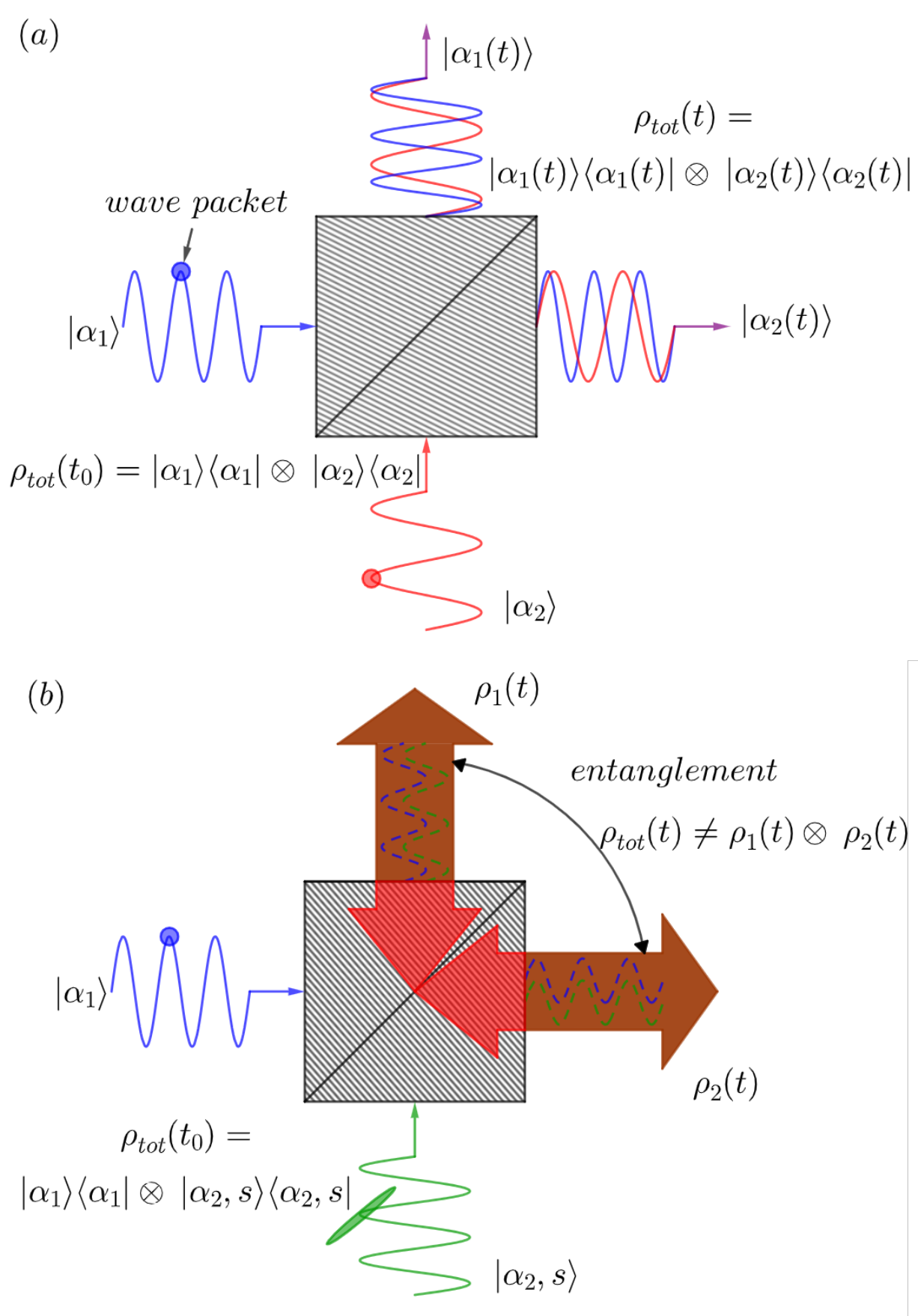}
\caption{(Colour online)  Schematic plots of (a) classical wave propagation given by coherent states and (b) quantum waves of entangling two 
modes through the beam splitting, where $\alpha_1(t) = u(t,t_0)\alpha_1+v_0(t,t_0)\alpha_2$ as shown by Eq.~(\ref{rdmps}), and $\rho_1(t)$ 
is given by Eq.~(\ref{rho_1(t) sol}). The detailed dynamics is fully solved from the deterministic quantum mechanics.}
\label{fig3}
\end{figure}

Moreover, if we let $s=0$ and $\alpha_1$ or $\alpha_2 =0$, namely one mode is initially in a coherent state and 
the other mode is in vacuum, then the dynamical evolution of both modes follow precisely the classical dynamics of two 
light waves passing through the beam splitting. The reduced density matrix operator of mode 1 is given by the pure state of 
Eq.~(\ref{rdmps}) with $\alpha_1(t) = v_0(t,t_0)\alpha_2$ or $\alpha_1(t)=u(t,t_0)\alpha_1$, where $u(t,t_0)$ and 
$v_0(t,t_0)$ are given in Eq.~(\ref{s_negf}). In Fig.~\ref{fig3}, we schematically plot the main difference of classical light 
wave propagation (the coherent state propagation) and quantum light wave propagation through the beam splitting.
The output light waves in the schematic plot of Fig.~\ref{fig3}(a) are called classical waves because they are characterized by the 
electromagnetic field eigenstates of the Glauber coherent states $|\alpha_1(t)\rangle$ and $|\alpha_2(t)\rangle$, even 
though they are fully solved from quantum mechanics. Figure \ref{fig3}(b) contains the quantum dynamics (entanglement 
dynamics) that classical processes cannot possess, because the causality of the dynamical evolution of each mode is broken
due to the squeezing effect.  In this case, one is unable to track the motions of waves or photons for each individual 
mode, even though the quantum dynamical evolution of the total state of the coupled two modes is deterministic.  
Figure \ref{fig3} shows the significant difference between classical-like wave and quantum wave through the same beam splittings.

\subsection{Comparison with other open quantum system approaches: coarse-graining vis fine-graining}
In the theoretical study of open quantum systems, the influence functional and other techniques have been extensively employed 
to derive the reduced dynamics of a subsystem \cite{Weissbook,Breuer2002}. The resulting reduced density matrix generally 
corresponds to a mixed state, signifying the emergence of non-causal statistical behavior apparently to open system dynamics. 
Nevertheless, most previous formulations, whether exact or approximate, lead to master equations that appear explicitly 
causal, often expressed in a time-convolutionless form. This superficial causality stands in contrast to the inherently 
non-causal nature of the reduced quantum dynamics itself, thereby obscuring the physical origin of statistical probability 
in open quantum dynamics and in quantum mechanics. Clarifying this conceptual inconsistency is therefore of fundamental importance. 
The present work addresses this issue directly by demonstrating how the causality is internally broken in the exact reduced dynamics
and thereby provides a natural and self-consistent explanation for the emergence of probability and mixed-state behavior in quantum mechanics.

%First, we may classify various techniques used in the studies of open quantum dynamics into three levels of coarse-grained. 
In most textbooks and common treatments of open quantum systems, coarse-graining generally refers to the approximate 
averaging over the fine details of the environment \cite{Breuer2002}. Typical examples include the Born-Markov master 
equation derived via perturbative expansion and the Lindblad master equation formulated within the semigroup framework. 
Both rely on the Markovian approximation, in which a non-causal statistical ensemble average is performed  
at every step of the reduced dynamics evolution. Other stochastic formulations fall into the same category. For instance, 
the stochastic Hubbard-Stratonovich decoupling proposed by Shao \cite{Shao2004} employs a Hubbard-Stratonovich 
transformation to recast the system–bath coupling into two subsystems driven by a common classical white-noise field. 
The use of white noise itself constitutes a Markovian approximation, representing a non-causal ensemble averaging process. 
%Consequently, the Markovian assumption is effectively imposed at every stage of the reduced dynamics for both the system and the bath. 
The resulting master equation is therefore also a Markovian-approximated one. In this sense, Markovian 
master equations that appear formally causal do not, in fact, reflect genuine dynamical causality at microscopic level.

%The influence functional \cite{Feynman1963} is also often called as a coarse-graining, but it is only in the mathematical sense 
%(defined by the trace over the environment $\rho_{\text{sys}}(t)=\Tr_{\text{env}} [\rho_{\text{tot}}(t)]$). 
Unlike approximate coarse-graining discussed above, the influence functional is exact and fully dynamical. It does not lose any correlations 
between the system and the environment. The environment effects are fully encoded in the nonlocal influence kernels \cite{Feynman1963,Leggett1983,GSI1988,Weissbook}, which carries all the memory and correlations 
between the system and environment. 
%Perhaps it is also worth emphasizing that the 
%conventional path integral found in most of quantum mechanics textbooks is simply described as a sum over all possible paths 
%connecting initial and final configurations.  However, the path integral expression of the unitary time-evolution operator can 
%be rigorously derived using either the identity resolution of the position states (path integral formulation in the position representation) 
%or the identity resolution of coherent states (path integral formulation in the coherent-state representation), where no a priori statistical 
%weights is assumed. The detailed derivations can be found in the textbooks by Feynman and Hibbs \cite{Feynman_Hibbs1965}, 
%Schulman \cite{Schulman1981} and Faddeev and Slavnov \cite{Faddeev1980}.  The coherent-state path-integral formalism is also 
%the standard functional formulation of quantum field theory \cite{Faddeev1980}. 
Technically, the time-nonlocal influence kernels %obtained by integrating exactly over all the environment degrees of freedom 
contains four different parts 
that modify the classical action of the reduced dynamics. The first two parts represent respectively the forward and backward 
propagations through the system-environment coupling in path integrals, one part comes from a single integral at the end 
point (at time $t$), which mixes the forward and backward propagations in all the early times. The last part arises from 
another single integral at the starting point (at time $t_0$) that mixes the forward and backward propagations in all the latter times
through the initial environment state.  This is the precise procedure how we derive the influence functional of Eqs.~(\ref{IFs})-(\ref{IFes}).
It also gives the physical picture of the closed time-path Green's function formulation for the reduced dynamics in open quantum systems, similar to
closed time-path Green's function formulation for the nonequilibrium dynamics in many-body system and quantum field theory 
\cite{CSHY1985,RS1986,WMZ1992}. This closed time-path picture for open quantum systems was 
well demonstrated through the exact influence functional in the position representation \cite{Feynman1963,Leggett1983,GSI1988,Weissbook} 
as well as in our coherent-state representation \cite{TZ2008,Jin_Tu2010,LZ2012,LZ2018,Zhang2019}. 

It is particularly important to notice that only the last part in the nonlocal influence kernels, arising from the single integral at the starting 
time $t_0$, causes the causality breaking in the reduced dynamics evolution. This is demonstrated explicitly in our stationary 
path equations of motion Eq.~(\ref{eom-1}) for the reduced system and the equations of motion Eq.~(\ref{eom-evo}) for the reduced nonequilibrium Green's 
functions. Thus, the reduced density matrix is fully determined by the reduced nonequilibrium Green's functions in the bilinear systems, as
we have also shown in our previous works \cite{TZ2008,Jin_Tu2010,LZ2012,LZ2018,Zhang2019,HZ2020,HZ2022b,HZ2022}.
We call this exact reduced dynamics as a {\it fine-graining} solution.  
The influence functional itself does not show you how the detailed reduced dynamics behaves, as we have pointed out in the 
end of Sec.~III.A. The deeper understanding of the reduced quantum dynamics lies on the {\it fine-graining}, 
from which the loss of deterministic predictability and the emergence of statistical probability in open quantum dynamics can be unveiled.  

As we have also pointed out in the end of Sec.~III.B, the previous influence functional applications (including the original work of 
Feynman and Vernon \cite{Feynman1963}) start with a thermal bath, which brings the statistical ensembles into the reduced 
subsystem dynamics from the starting conditions. Explicitly, if we set the initial bath at zero 
temperature ($T=0$), the initial bath state becomes a vacuum state, no thermal fluctuations exist. Then the nonlocal influence kernel 
relating to the initial environment state vanishes immediately, as shown by Eq.~(\ref{sbcc}). Thus, the open quantum system 
becomes a pure dissipative system which cannot make the statistical ensembles emerges in the reduced subsystem dynamics.
Maintaining fluctuation-dissipation relationship becomes crucially important in the usual study of open quantum systems. 

In this work, we set up an extremely simple environment, consisting of only one mode. With such a single-mode environment, the energy 
constantly oscillates between the system and environment, as shown in the exact master equation Eq.~(\ref{merho1}), so that it has no dissipation. 
Furthermore, we initialize the environment in a squeezing coherent state, ensuring the absence of thermal fluctuations (and therefore, 
no statistical ensemble) from the beginning. Our exact solution (fine-graining reduced dynamics) shows that the pure quantumness of the initial squeezing state
breaks the causality of the reduced dynamical evolution, thereby unambiguously demonstrating the dynamical process that 
leads to the loss of deterministic 
evolution and the emergence of statistical probability in quantum mechanics. This has nothing to do with dissipation and thermal fluctuations.

Our exact master equation Eq.~(\ref{merho1}) is also time-convolutionless one, namely a standard causal 
equation of motion. The system contains neither dissipation nor thermal fluctuations, how the solution of the reduced quantum dynamics 
is given by a mixed state, demonstrating the emergence of non-causal statistical behavior? The answer to this question is hidden in these 
time-dependent parameters in the master equation Eq.~(\ref{merho1}). These time-dependent parameters are given in 
Eq.~(\ref{mepara}). They are determined by the reduced retarded Green's function $u(t,t_0)$ and the reduced correlated Green's functions 
$v_1(\tau,t)$ and $v_2(\tau,t)$. These reduced nonequilibrium Green's function satisfies the time-convolution equations of motion 
Eq.~(\ref{eom-evo}) which are derived from the explicit noncausal and  time-convolution  equation of motion for the stationary 
paths of the propagating functional Eq.~(\ref{propagating}). These relations
was also explored in our early study of nanoelectronic nonequilibrium dynamics, collaborated with Prof. Yijing Yan \cite{Jin_Tu2010}. 

Two reduced correlated Green's functions $v_1(\tau,t)$ and $v_2(\tau,t)$ are crucially important, representing the non-causal evolution arisen from 
the quantumness of the squeezing state. Note that for fermionic open systems, fermionic vacuum state contains pure quantumness. 
The reduced correlated Green's functions also do not vanish even for the zero temperature fermionic bath  \cite{Jin_Tu2010}, which is 
significantly different from a bosonic bath.  In other words, we show that the internal causality breaking is purely induced by quantumness. 
Because of the internal causality breaking in the exact master equation, special attentions 
need to be paid when one numerically solve the master equation. In every time step of calculating the reduced density matrix, 
one must repeatedly calculate these time-dependent parameters from the very initial time $t_0$, because the different choice of a latter
time $t$ will change the past evolution of the stationary paths, a truly advanced effect as we shown in Fig.~\ref{fig1}. 
This gives the physical picture how to observe non-causality in the apparently causal exact master equation.

In conclusion, the microscopic non-causality is inherent in reduced quantum dynamics. However,  most approximations used 
to study open quantum systems, as well as the influence functional derived with a thermal bath, discard the physical 
picture for the emergence of statistical nature in reduced quantum dynamics. 
%It thereby discards the dynamical genesis of entanglement between the system and the environment.
The above analysis shows that these non-trivial results presented in this work is not due to our unique method, but rather 
from the elimination of various ambiguities, thereby deepening our understanding of the underly dynamics in open quantum 
dynamics and in quantum mechanics.  %This could be an revolutional breakthrough in the foundation of quantum mechanics.
%For clarity, we may broadly categorize research in open quantum systems  into three levels: 1) Approximate coarse-graining 
%(Markovian-type master equations) in most applications; 2) Exact coarse-graining (influence-functional approaches), and
%3) Fine-graining reduced dynamics (our previous works, nonequilibrium Green's function technique).

\iffalse
It may be also worth mentioning that 
in the literature, one has taken an one-photon state to demonstrate the delayed-choice gedanken experiment 
proposed by Wheeler \cite{Wheeler1978} in interferometric setups to test particle-wave duality, which uses the same 
picture as shown Fig.~\ref{fig3}(a) and is indeed classical. A true one-photon state, namely the Fock state 
$|n=1\rangle$ is pure quantum mechanical, there is no classical correspondence for a single photon state $|1\rangle$, its Wigner 
distribution has the negative values that cannot be represented by the processes of Fig.~\ref{fig3}(a).  The out-coming state coupled
with another mode through a beam splitting forms an entanglement state as well, and therefore it has the picture of Fig.~\ref{fig3}(b) 
rather than Fig.~\ref{fig3}(a). As a result, Wheeler's idea of the delayed-selected gedanken experiment using a single photon passing 
through a beam splitting to demonstrate particle-wave duality should be re-examined. In fact,
the detailed physical picture of a "truly" one-photon state evolution coupling with another mode or multimode systems is nontrivial. 
We will leave this investigation of single photon state evolution in another publication. %\cite{Yang24a}. 
\fi

\section{Discussions and Perspectives}
\label{concl}
In this paper, we study the physical underpinning of the entanglement emergence from the quantum evolution of a 
coupled two-mode system, a very simple system that can be easily implemented and demonstrated experimentally. 
The coupling is linear, so it does not inherently create entanglement between the two modes. 
We also start with separable initial pure states of the two modes, so that there are no entanglement and also no statistics 
to begin with. In particular, we set one mode initially in a Glauber coherent state which is a wave packet with minimum
uncertainty ($\Delta x = \Delta p$ and $ \Delta x \Delta p= \hbar/2 $) which is one-to-one corresponding to a classical particle 
in a harmonic potential.  Thus, by looking at 
the quantum evolution of this mode governed by the coupled Hamiltonian Eq.~(\ref{Htot}), we have shown in what conditions
the evolution of the Glauber coherent state can give rise to exactly the same classical dynamics, and under what circumstances 
the Glauber coherent state will tune to be a mixed state so that the two modes evolve into an entangled state. 

With such an investigation, we provides the way to explore the essence and the origin of the entanglement in quantum realm. We find that
the emergence of entanglement is accompanied with the internal causality breaking, and both stem from
the quantumness containing in the initial state (in this paper we used the squeezed coherent state) of another mode.
If another mode is also initially in a Glauber coherent state, then entanglement cannot emerge and causality does not
be violated in the dynamical evolution of each mode, the corresponding quantum dynamics gives rise precisely the 
same classical dynamics of the two coupled harmonic oscillators. This answers the question why entanglement emerges in the quantum 
world but cannot occur in classical deterministic physics.

The system and the setting studied in this paper are simple but quite unique. This is because the Glauber coherent states (wave packets 
with minimal Heisenberg uncertainty) are the only quantum states that behave as classical particles that can evolve exactly along their classical 
trajectories in a harmonic potential. It provides a unique way to distinguish the differences between quantum and classical dynamics 
in terms of the same language. It also demonstrates explicitly the quantum properties that are not present in classical physics. 
In the seminal EPR paper \cite{einstein1935can}, Einstein {\it et al.}~proposed the thought experiment of using 
a plane-wave wavefunction to describe the entanglement phenomenon between two particles at a distant that the interaction between them can be ignored. 
It is a thought experiment because a plane-wave wavefunction is quantum mechanically unphysical \cite{Tannoudji1991}, 
the corresponding probability is equally distributed in the space and the overall probability is divergent so 
experimental measurements for such physical particles are infeasible. 
David Bohm reformulated the EPR experiment later by considering a pair of entangled spin-$1/2$ 
particles that can be measured \cite{Bohm1951}. But a spin system (or more generally, fermionic systems)  is intrinsically a quantum system 
that differs from classical systems. In fact, any system containing only two or a finite number of states has no direct correspondence with 
a classical system. It is difficult, if not impossible, to find other systems and setups such that whose quantum dynamics can exactly reproduce the 
corresponding classical dynamics  on the one hand \cite{ZhangPR95,Zhang1990,Zhang1994}, and on the other hand, demonstrate these
quantum properties that cannot present in classical physics by simply changing their initial states.

Although the system and setups in this paper are very simple and quite unique, the conclusion we obtained is 
actually general and non-trivial.  The exact solution of the reduced density matrix operator $\rho_1(t)$ of mode 1 solved from the deterministic 
evolution equation of the two-mode coupling system can be straightforwardly extended to many-mode coupling systems, including 
continuously distributed infinite number of modes. As long as the coupling is bilinear, and the initial states of  
other modes are not all in coherent states, entanglement inevitably emerges. The emergence of entanglement  
is accompanied by the internal causality breaking in the dynamical evolution of each mode, as manifested through 
the same equations of motion of Eq.~(\ref{eom-1}), where it only needs to slightly change the two-time correlations of Eq.~(\ref{ttcf})
to a sum over all other modes when more modes are included, as shown by Eq.~(\ref{ttcf_ext}). This formulation 
has been shown in our general investigations to open quantum system dynamics \cite{LZ2012,LZ2018,HZ2022b}. 

The further extension to many-fermionic systems and topological dissipative systems is also straightforward,
by changing the complex variables to the Grassmann variables in the same equations of motion for the stationary 
paths with some sign changes caused by the anti-exchange property of Grassmann variables \cite{TZ2008,Jin_Tu2010,HZ2020}.
Thus, emergence of entanglement accompanying with the internal causality breaking is manifested in the same way 
for fermionic systems and topological systems. 
 In fact, this is a general consequence for any composite systems, the emergence of  entanglement between 
two or more subsystems are accompanied by causality violation in the reduced dynamical evolution 
of these subsystems, except for linear coupling  bosonic systems with all particles being initially in wave packets with 
minimum Heisenberg uncertainty whose dynamics are causally classical.  
As we have shown, the final solutions from the exact master equation obtained in this work 
are determined by Eq.~(\ref{eom-evo}), which corresponds to the reduced nonequilibrium Green's functions
that can be easily and straightforwardly extended to any many-body quantum systems and quantum field theory, where different interactions will 
result in different and more complicated two-time correlations between different subsystems than that given by Eq.~(\ref{ttcf_ext}).
An analytical solution of the reduced nonequilibrium Green's functions, as given by Eq.~(\ref{s_negf}), usually cannot be obtained 
beyond the bilinear systems \cite{Zhang2019,HZ2022}, but they can be calculated approximately with the standard many-body
theory \cite{CSHY1985,RS1986,WMZ1992}. Thus, the internal causality breaking found in the bilinear systems is an nnherent 
quantum property to the reduced quantum dynamical evolution in any interacting system.

As an extension, the rigorous and exact solution obtained from the simple composite system in this work shows that   
causality violation only occurs for reduced subsystem dynamics. This is why we called it as an internal causality breaking. While, the evolution 
of the whole systems (i.e., isolated systems) follows the deterministic Schr\"{o}dinger equation.  It is widely true that the quantum 
dynamical evolution of various constitutes in quantum systems all breaks internally the causality.  Schr\"{o}dinger equation is the quantum equation 
of motion for isolated systems only. Any constitute in a physical system cannot be treated as an isolated system, even for the single 
electron in hydrogen atom. As an illustration, the electron in atoms must interact with nucleus to form the bound states and it also interacts with photons 
to give the transitions between different states. Therefore,  electron and its movement is only a part 
of the quantum dynamics of atoms, molecules and solids. 
Our solution indicates that the causality in the quantum evolution of every electron should be internally 
broken, even though the corresponding electron movement would be more complicated than that in the example we given in this paper and cannot be 
simply determined by stationary paths alone for the coherent-state path integrals. 

More importantly, the internal causality breaking naturally leads to a statistical description for quantum measurements, because when the system 
is measured, it is usually no longer an isolated system. %\cite{Neumann_Landau_1927}.  
Consequently, it is hard to see precisely the movement of single electron in atoms, molecules and solids, even though some 
dynamical effects of single electron  have been observed by means of the attosecond spectroscopy \cite{Krausz2010,LHuillier2017,Krausz2009,Agostini2014}.  
It should not be possible to track the electron (and any particle) movement 
when it transits between different states, not because of the very short timescales but due to the internal causality breaking 
for the dynamics of various particles in quantum realm.   The lack of causality is the nature of statistics. 
This reveals the long-standing mystery why quantum mechanics holds a probabilistic 
interpretation. It is the internal causality breaking in quantum dynamics makes the measurement results become probabilistic. 

In fact, to describe the dynamical process of  quantum measurement, one should consider the measured system 
plus the measuring apparatus together as a composite system, as was originally proposed by von Neumann \cite{vonNeumann1927,vonNeumann1932}. 
With this extension, the composite system of the observed system plus the measuring apparatus forms a closed system and
can follow the unitary Schr\"{o}dinger evolution. While the measured system or the measuring apparatus alone is no longer a closed 
system and does not follow the unitary Schr\"{o}dinger evolution. 
As a result of the interaction between the system and the apparatus, their joint state is still a pure state but each of the subsystems is 
in a mixture. Nowadays, this becomes well-known as a natural consequence of the entanglement between the observed system and 
the measuring apparatus, but it was indeed discovered by Landau \cite{Landau1927} and also was extensively discussed by von Neumann 
\cite{vonNeumann1932}  in the very early development of quantum mechanics in 1927.   

Apply our two-mode 
system to measurement process, one mode can be considered as the measured system and another mode can be regarded as
the measuring apparatus. The equation of the motion Eq.~(\ref{merho1}) provides the exact dynamics of such measurement process
and shows how the unitarity of the measured system is broken down during the measurement.
The corresponding exact solution of the reduced density matrix $\rho_1(t)$ determined by the non-causal stationary paths of 
Eqs.~(\ref{eom-1}) and (\ref{z1 relation}) presented in this work unveils the origin of statistical nature in quantum 
mechanics through the measurements. That is, the internal causality breaking in the evolution of the measured system 
naturally makes measured quantum states become mixed states such that measurement results occur only probabilistic.  This
conclusion is reached by solving the deterministic Schr\"{o}dinger equation for the composite system model of the measured system plus the measuring
apparatus, without requiring a prior probabilistic interpretation or a statistical ensemble assumption regarding the initial wavefunction of the system.  
This solution also dynamically answers the question why the measurement processes in quantum mechanics is non-causal, 
as was originally discussed by von Neumann \cite{vonNeumann1932}. In this sense, the non-causal measurement processes are 
just a particular consequence of the internal causality breaking in the reduced dynamics of open quantum systems. 

As we have also shown in this simple two mode system, the internal causality breaking is mathematically defined and manifested in the  
equations of motion only for the reduced dynamics, namely the future dynamics change the current dynamics of the reduced subsystems.  
Apply this conclusion to measurement processes, it indicates that the future dynamics of the measured system can 
change the current dynamics of the measured system, as a consequence of the mixing forward and backward propagating dynamics
of the the measured system through the interaction with the measuring apparatus. Here the forward and backward propagation dynamics 
are given by replacing the original Schrödinger equations of the wave function with the von Neumann evolution equations 
of the density matrix for the closed system, representing the unitary evolution of the closed system, see Eq.~(\ref{ted}), that manifests the time symmetry
of quantum evolution, i.e. the reversibility. 
This time evolution symmetry or reversibility is broken down for the measured system by the interaction with the measuring apparatus when a 
measurement is performed, as we solved dynamically in this work with the simple two mode system. 

We may point out that the above dynamical description to quantum measurement processes should not be confused with the time-symmetric 
formulation of the quantum measurement proposed by Aharonov {\it et al.} \cite{ABL1964,APT2010}, in which additional forward and backward  
propagating waves are introduced to the interpretation of pre- and post-selected measurement ensembles, or as a conditional probabilistic interpretation, 
to describe weak values and retrodictive information flow, where the time symmetry of the pre- and post-selected ensembles 
does not represent the real dynamical evolution of quantum measurement processes.
A complete solution to the selective quantum measurement requires to include the interaction with the observer, namely the whole 
measurement process involves the dynamics of the measured system, the measuring apparatus and the observer, plus the interactions
between the system and the apparatus and also the interaction between the apparatus and the observer \cite{vonNeumann1932}. 
The solution to this larger composite system is still a major challenge, even though it might give a understanding to the dynamics of the 
post-selected measurement, or otherwise the dynamics of the wavefunction collapse, the long historical debate in quantum mechanics 
\cite{Zuerk2003,Nature2025}. 
The exact solution of the simplest composite system based on the first principles of quantum mechanics presented in this work 
could become a crucial step towards solving this challenge.
%In the Schr\"{o}dinger's formulation  of quantum mechanics, the measurement instruments are not included. The present work and results 
%can be considered as a dynamical realization of the quantum measurement in terms of the simplest system to show how the mixture and 
%probability emerges through the dynamical evolution of composite system of the system to be observed plus the measuring instruments.  
%Thus, the probability interpretation adding to the Schr\"{o}dinger's formulation can be naturally obtained
%Thus, this could provide a underlying equation of motion for the subsystems of simply composite systems, like Eq.~(\ref{merho1}), 
%to explore the dynamical process of quantum measurement, the problem that was originally proposed by von Neumann and Landau 
%in the very early development of quantum mechanics \cite{Neumann_Landau_1927}.

At last, one might ask how such internal causality breaking in quantum systems can be detected experimentally. To a certain extent, it 
requires to detect the advanced-time effect rather than the retarded-time effect in the dynamical evolution, which is far more than just 
a difficulty.  Although measuring the two-time correlations of the quadrature matrix in quantum optics is not a big problem,
state-of-the-art experimental setups are needed. The advanced experimental technologies, such as precision measurement or 
non-demolition measurements 
%\cite{Thorne1980,Haroche2007a,Haroche2007b} 
and the faster detection with attosecond spectroscopy 
%\cite{Krausz2009,Agostini2014} 
may  be useful tools. We leave this practical measurement problem to experimentalists. 
Nevertheless, we show that the internal causality breaking in quantum systems is 
an inherent property of quantum mechanics. 
It opens up a new avenue to explore many related fundamental physics problems that have not been fully understood so far, such as the foundation 
of thermalization in nature, the dynamical evolution of bio-systems, and even the origin and the evolution of our universe, etc. 
The dynamic processes in these fields are dominated with nonequilibrium in nature and cannot follow the principle of causality 
at the atomic level from the open system point of view. 

Furthermore, the internal causality breaking in finite quantum systems may also make a new challenge to the development of 
quantum technology.  As one knows, the main obstacle for the quantum technology comes from the inevitable 
decoherence effect.  From our previous research on decoherence theory \cite{TZ2008, Zhang2012, LZ2018, HZ2022b}, one would find 
that decoherence stems indeed from the same reason as the emergence of entanglement and statistics in quantum systems, 
i.e., the internal causality breaking. The same problem also exists for topological states (Majorana zero modes, for an example \cite{LZ2018,HZ2020}).    
Although noisy intermediate-scale quantum (NISQ) processors have been rapidly developed over the past few years, they may mainly be of 
practically useful for quantum simulations. Our finding of internal causality breaking in finite quantum systems raises a more fundamental 
question: whether unitary operations on individuals in a composite system is practically feasible.
In other words, performing internal unitary operations on a single qubit or a group of qubits in a large-scale qubit system is virtually 
as difficult as solving completely the decoherence problem, due to the various entanglements among qubits for subsequent operations 
on qubits \cite{Horache1996}.
Therefore, how to compatibly utilize the properties of entanglement and overcome the difficulties arisen from  internal causality breaking 
is a great challenge. We leave this issue open for further investigation.

%In conclusion, the dynamic quantum fluctuation plays a vital role in quantum entanglement. The dynamic quantum fluctuation 
%causes that the pure state become the mixed state such that there exist quantum entanglement between two subsystems, 
%even if the propagating function which consists of coherent state in the Hamiltonian (Eq.~(\ref{Htot})) has
%the classical properties. The dynamic quantum fluctuation also causes the contribution of the path breaking the 
%causality in the evolution of the subsystem, so there doesn't satisfy the causality in the subsystems. 
%This illustrate that quantum entanglement must not happen in the classical mechanics because particle 
%in the classical mechanics even doesn't have quantum fluctuation, and the causality is established in the subsystem in classical mechanics.

\ \\

\acknowledgments

WMZ would like to express his special gratitude to Prof.~Yijing Yan for the early collaboration on research in open quantum systems, 
particularly their joint work on developing a unified description of nonequilibrium quantum state dynamics by bridging the Feynman–Vernon 
influence functional method with the Schwinger–Keldysh nonequilibrium Green’s function technique for open quantum systems \cite{Jin_Tu2010}. He also warmly 
thanks Prof.~C.~Q. Geng and Prof.~X.~G. He for their generous hospitality during his visits to the Hangzhou Institute for Advanced Study, 
University of the Chinese Academy of Sciences, and the Tsung-Dao Lee Institute of Shanghai Jiao Tong University, where part of this work 
was carried out.
This work is supported by National Science and Technology Council of Taiwan, Republic of China, under Contract
No.~MOST-111-2811-M-006-014-MY3.

\end{document}